\documentclass[aps,epsf,preprint]{revtex4}
\usepackage{natbib}
\usepackage{latexsym}
\usepackage{amsfonts}
\usepackage[dvips]{graphicx}
\usepackage{color}
\usepackage{float}

\begin{document}


\title{Modeling Amphiphilic Solutes in a Jagla Solvent}

\author{Zhiqiang~Su,$^1$ Sergey~V.~Buldyrev,$^{1,2}$
  Pablo~G.~Debenedetti,$^3$ Peter~J.~Rossky,$^4$ and
  H.~Eugene~Stanley$^1$}

\affiliation{$^1$Center for Polymer Studies and Department of Physics,
  Boston University,~Boston, MA 02215\\ 
$^2$ Department of Physics,~Yeshiva University, 500 West 185th
  Street,~New York, NY 10033 
\\ $^3$Department of Chemical and Biological Engineering, Princeton University,
Princeton, NJ, 08544\\ 
$^4$Department of Chemistry and Biochemistry, College of Natural Science,
The University of Texas at Austin, Austin, TX, 78712} 

\date{6 Sept. 2011 --- sbdrs6sept.tex -- JCP submitted}

\begin{abstract}

Methanol is an amphiphilic solute whose aqueous solutions exhibit distinctive physical
properties. The volume change upon mixing, for example, is negative across the entire 
composition range, indicating strong association. We explore the
corresponding behavior of a Jagla solvent, which has been previously
shown to exhibit many of the anomalous properties of water. We consider two
models of an amphiphilic solute: (i) a ``dimer" model, which consists of
one hydrophobic hard sphere linked to a Jagla particle with a permanent bond, and
(ii) a ``monomer" model, which is a limiting case of the dimer, formed
by concentrically overlapping a hard sphere and a Jagla particle. Using
discrete molecular dynamics, we calculate the thermodynamic properties
of the resulting solutions. We systematically vary the set of
parameters of the dimer and monomer models and find that one can readily
reproduce the experimental behavior of the excess volume of the
methanol-water system as a function of methanol volume fraction. We
compare the pressure and temperature dependence of the
excess volume and the excess enthalpy of both models with
experimental data on methanol-water solutions and find qualitative
agreement in most cases. We also investigate the solute effect on the temperature of maximum density and find that
the effect of concentration is orders of magnitude stronger than measured experimentally.

\end{abstract}

\maketitle

\section{INTRODUCTION}

Aqueous solutions of alcohols have attracted both
experimental and theoretical attention on account of their ubiquity
and importance in the medical, personal care, transportation 
(e.g., antifreeze, fuels), and food industries, among others
\cite{AkerlofJAC1932,GibsonJAC1935,FrankJCP1945,LamaJCED1965,BensonJSC1980,KubotaIJT1987,TomaszkiewiczTA1986II,TomaszkiewiczTA1986III,XiaoJCT1997,Mcglashan1976,SatoJCP2000}.
Methanol is a simple example of an amphiphilic organic
solute, and its aqueous solutions exhibit many interesting nonidealities. 
Attaining a good understanding of this simple case is therefore a natural 
starting point for studies of more complex solutes in water, such as
higher alcohols or proteins.

Because of the increasing availability of expanded computing power,
simulations have become an important research tool in studying aqueous methanol solutions
\cite{MauroJCP1990,HidekiJCP1992,EickJCP2003,Diego2006,Imre2008,YangJCC2007}.
The optimized potential for liquid simulation (OPLS)
\cite{JorgensenJPC1986} is frequently used to model alcohol
molecules. To represent water as a solvent, the SPC/E
\cite{BerendsenJPC1987}, TIP3P \cite{JorgensenJCP1983}, TIP4P
\cite{JorgensenMP1985}, and TIP5P \cite{MahoneyJCP2000} models are
frequently used. On the other hand, many thermodynamic properties of
water can be reproduced using soft-core spherically symmetric
potentials, one of the most important of which is the Jagla model
\cite{JaglaPRE2001}. The Jagla potential has a hard core and a linear
repulsive ramp, and contains two characteristic length scales: a hard
core $a$ and a soft core $b$. For a range of parameters, the Jagla model
exhibits a water-like \cite{ErringtonNature2001} cascade of structural, transport, and
thermodynamic anomalies
\cite{JaglaPRE2001,YanPRL2005,LimeiPNAS2005,LimeiPRE2006}.
Buldyrev et al. in 2007 \cite{SergeyPNAS2007} found that the Jagla
solvent exhibits key water-like characteristics with respect to hydrophobic
hydration, suggesting that the water-like characteristics of the Jagla
solvent extend beyond the pure fluid. Here, we explore this analogy
further, considering the properties of solutions of amphipathic solutes.

We focus on the properties of the excess volume and the excess enthalpy,
and on how amphiphilic solutes affect the temperature of maximum density
($T_{\rm MD}$) of a solution.  At $T=298$K and $P=0.1$MPa
the excess volume and the excess enthalpy are negative across the entire
range of methanol concentrations (both quantities will be 
defined rigorously below). The strongest effect occurs at a methanol volume
fraction of $\varphi_{\rm max}=59.7$\%, at which the negative excess volume deviates from additivity by $-3.57$\%
\cite{Mcglashan1976}.  As $T$ increases and $P$ decreases, the excess
volume becomes more negative and the extremal point shifts (see
Fig.~\ref{temperatureexperiment} and Fig.~\ref{pressureexperiment}).
Most solutes tend to suppress water's non-idealities, and hence they lower
the $T_{\rm MD}$. Some amphiphilic solutes behave differently: ethanol and t-butanol
have a marked non-monotonic effect (whereby the $T_{\rm MD}$ of the solution first
increases with respect to that of water upon solute addition, but decreases for
more concentrated solutions), and methanol has a very mild non-monotonic effect
\cite{WadaBCSJ1962}.

As a first step , in this paper, our goal is to model a simple
amphiphilic solution that mimics the properties of the methanol-water
system. We use a hard sphere with parameterized diameter $a_{\rm M}$
to model the hydrophobic methyl group, and a Jagla particle to model the
hydrophilic hydroxyl group. We link the hard sphere and the Jagla
particle using a bond of adjustable length to model the amphiphilic
solutes.

In Section II of this paper we describe in detail our models and
simulation methods. In Section III we list and analyze the simulation
results, and in its four subsections we discuss the different parameter
effects, the temperature and pressure dependence of the excess volume, the
behavior of the excess enthalpy, and how the solute concentration
affects the temperature of maximum density. In Section IV we list our main
conclusions.

\section{MODEL and METHODS}

\subsection{Dimer model for amphiphilic solutes} 

To model the simple amphiphilic solute methanol, we separate CH$_3$OH into
the methyl (CH$_3$) and the hydroxyl (OH) groups. We model CH$_3$ as a
hard sphere and OH as a Jagla particle.  There is one bond between the
hard sphere and the Jagla particle. We call this model the dimer model.
To model the solvent, H$_2$O, we use the Jagla
particle. The following interactions are included: 
there is a Jagla potential between Jagla solvent particles, a
Jagla potential between the Jagla solvent and a dimer's Jagla particle, and a
Jagla potential between two dimer's Jagla particles, all of which are denoted by 
$U_{\rm JJ}(r)$ [Fig.~\ref{model}(a)]. 
The interaction between the hard spheres is modeled by a
hard-core potential $U_{\rm HH}(r)$, and the interaction between the
hard spheres and the Jagla particles is modeled by a hard core potential
$U_{\rm JH}(r)$. We model the covalent bond with a narrow square well
potential bounded by two hard walls $U_{\rm bond}(r)$. The interaction
potentials are
\begin{equation}
U_{\rm JJ}(r) = \left\{ \begin{array}{ll}
\infty & r< a \\
-U_{o}+\frac{(U_{o}+U_{R})(b-r)}{b-a} & a<r<b \\
-U_{o}\frac{c-r}{c-b} & b<r<c \\
0 & r>c
\end{array} \right. ,
\end{equation}
where $a$ is the hard core diameter, $b=1.72a$ is the soft core
diameter, $c=3a$ is the range of attractive potential, $U_{o}$ is the
maximum attractive energy and $U_{R}=3.56U_{o}$ is the maximum repulsive
energy, and
\begin{equation}
U_{HH}(r) = \left\{ \begin{array}{ll}
\infty & r< a_{\rm M} \\
0 & r> a_{\rm M}
\end{array} \right. ,
\end{equation}
where $a_{\rm M}$ is the diameter of the hard sphere, and
\begin{equation}
U_{JH}(r) = \left\{ \begin{array}{ll}
\infty & r< a_{\rm JM} \\
0 & r> a_{\rm JM}
\end{array} \right. 
\end{equation}
where $a_{\rm JM}=\frac{a_{\rm M}+a}{2}$ and
\begin{equation}
U_{\rm bond}(r) = \left\{ \begin{array}{ll}
\infty & r < \zeta-\frac{\delta}{2}\\
0& \zeta-\frac{\delta}{2} < r < \zeta +\frac{\delta}{2} \\
\infty & r > \zeta +\frac{\delta}{2}
\end{array} \right. ,
\end{equation}
The hard core diameter $a_{\rm M}$ and the average length of the covalent
bond $\zeta$ are used as adjustable parameters in order to achieve agreement between the excess volume of
the model solution and the experimental results of methanol-water
solutions at ambient conditions. In all our simulations, we use the same
set of Jagla potential parameters, $b=1.72a$, $c=3a$, and
$U_{R}=3.56U_{o}$ \cite{LimeiPRE2006}. We use reduced
units in terms of length $a$, energy $U_{o}$, and particle mass $m$. For
temperature we use units of $U_{o}/k_{B}$; for pressure we use units of
$U_{o}/a^3$; and for volume we use units of $a^3$.

\subsection{Monomer model for amphiphilic solutes}

\begin{figure}[H]
\includegraphics[width=0.5\textwidth]{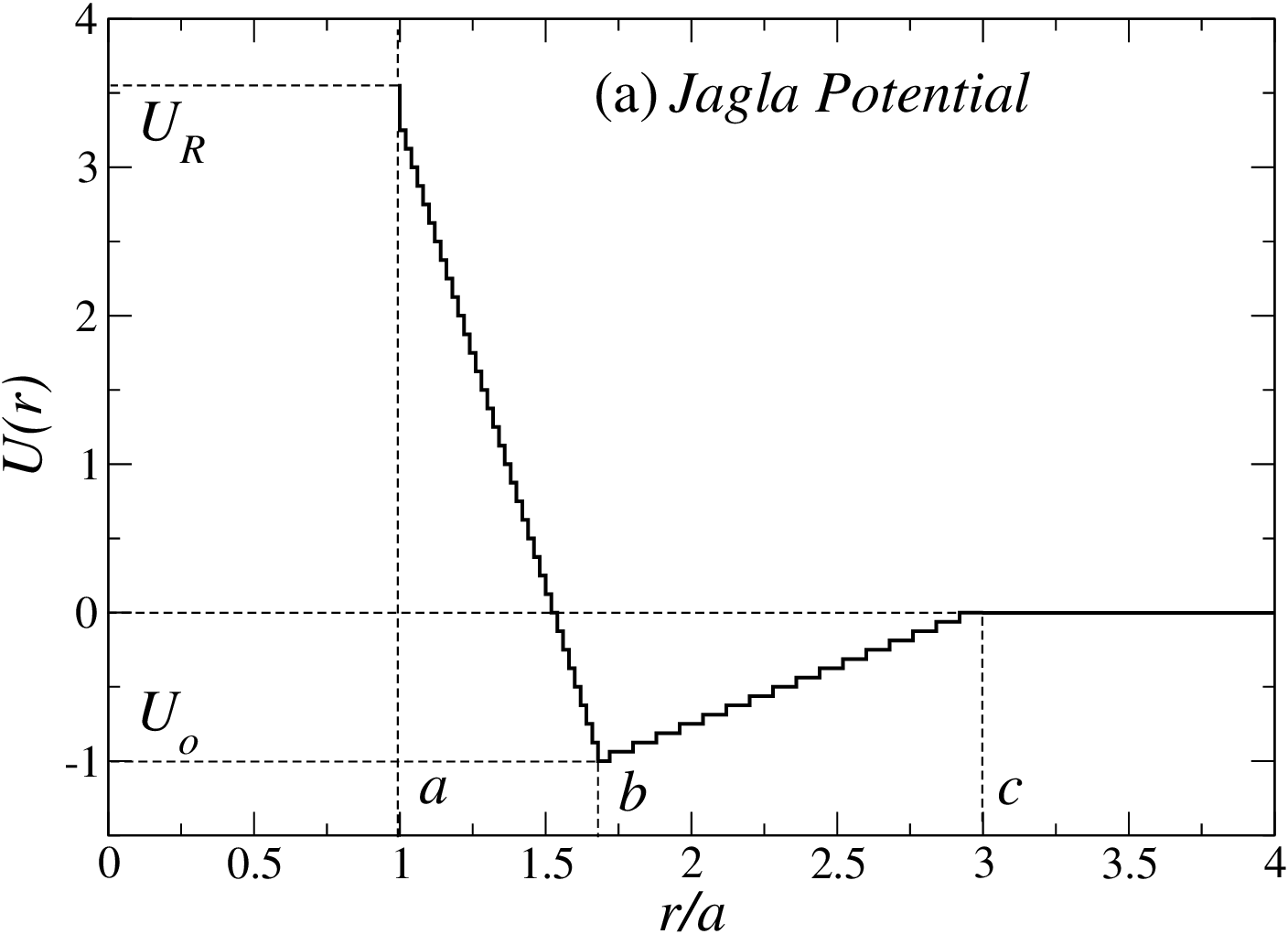} \\
\includegraphics[width=0.5\textwidth]{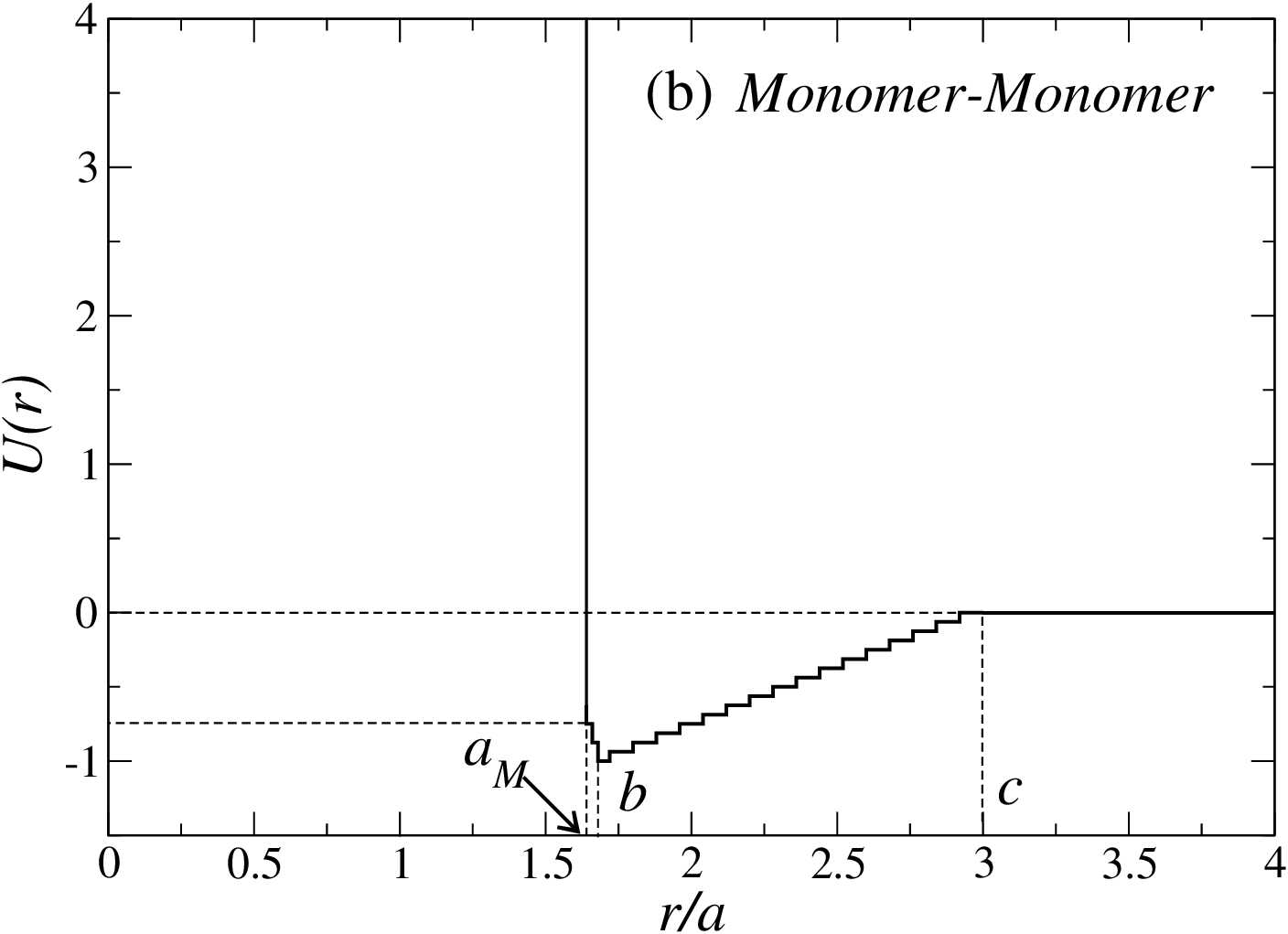} \\
\includegraphics[width=0.5\textwidth]{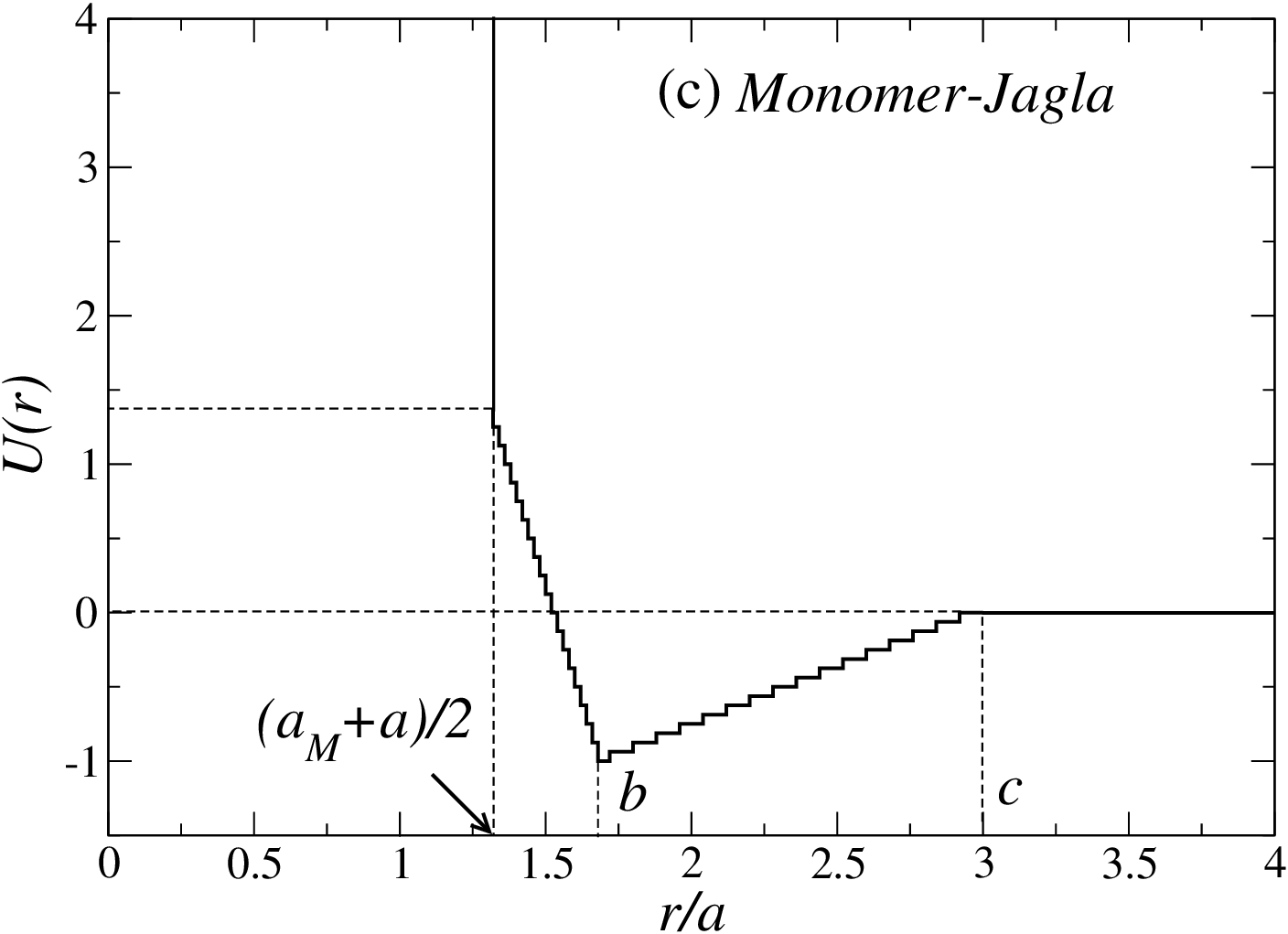}
\caption{Sketch of the interaction potentials in our simulation. (a) The
  spherically symmetric ``two-scale'' Jagla ramp potential.  The two
  length scales are the hard core diameter $r=a$, and the soft core
  diameter $r=b$. We study the case $U_{R}=3.56U_{0}$, $b=1.72a$ and a
  long range cutoff $c=3a$. (b) The interaction potential between two
  monomers, with contact distance $a_{\rm M}$.  (c) The interaction
  potential between the Jagla particle and the monomer, contacting at
  $(a+a_{\rm M})/2$.}
\label{model}
\end{figure}

The best agreement between the dimer model and the experimental data
occurs when bonds are short (for a more detailed discussion, see
Sec.~II), i.e., the hard sphere that models the methyl group and the
Jagla particle that models the hydroxyl group almost overlap. For this
reason we also consider a methanol model in which the overlap is
complete, and the bond length vanishes.  This leads to a spherically
symmetric potential that superimposes the Jagla particle and the hard
sphere. In this ``monomer'' model we introduce the interaction
potentials between ``methyl'' monomers, $U_{\rm MM}(r)$ [Fig.~\ref{model}(b)], between monomer
and Jagla particle, $U_{\rm JM}(r)$ [Fig.~\ref{model}(c)], and between Jagla particles $U_{\rm
  JJ}(r)$ [Fig.~\ref{model}(a)]. In this case the interaction formulae are
\begin{equation}
U_{\rm MM}(r) = \left\{ \begin{array}{ll}
\infty & r< a_{\rm M} \\
 U_{\rm JJ}(r) & r > a_{\rm M}
\end{array} \right. ,
\end{equation}
and
\begin{equation}
U_{\rm JM}(r) = \left\{ \begin{array}{ll}
\infty & r< a_{\rm JM} \\
U_{\rm JJ}(r) & r > a_{\rm JM}
\end{array} \right. ,
\end{equation}
where $U_{\rm JJ}(r)$ is
defined by Eq.~(1).

\subsection{Simulation details and analysis methods}

For our simulations we use the discrete molecular dynamics (DMD) algorithm. 
With DMD we approximate a continuous potential by a discrete
potential made up of a series of steps. We use the same scheme as in
Ref.~\cite{LimeiPRE2006}. Our simulation consists of a fixed number $N=2000$ particles
in a cubic box with periodic boundaries.  We denote the solute mole
fraction by $x$. Since the dimer contains two particles and the monomer one, 
the number of solute molecules $N_{s}(x)$ is
\begin{equation}
N_{s}(x) = \frac{Nx}{x + 1}
\end{equation}
in the dimer system, and 
\begin{equation}
N_{s}(x) = Nx
\end{equation}
in the monomer system.  The number of the solvent particles $N_{J}(x)$
is
\begin{equation}
N_{J}(x) = \frac{N(1-x)}{x + 1}
\end{equation}
in the dimer system, and
\begin{equation}
N_{J}(x) = N(1-x)
\end{equation}
in the monomer system.  The total number of molecules in the system
$N_{T}(x)$ is
\begin{equation}
N_{T}(x) = \frac{N}{x + 1}
\end{equation}
in the dimer model, and
\begin{equation}
N_{T}(x) = N
\end{equation}
in the monomer model.  The volume occupied by $N_{J}(x)$ pure Jagla solvent particles before mixing, $V_{J}(x)$,
and the volume occupied by $N_{s}(x)$ pure solute molecules before mixing, $V_{s}(x)$, at the given temperature
and pressure, are given by 
\begin{equation}
V_{J}(x) =\frac{N_{J}(x)}{N_{J}(0)}V_{\rm mix}(0)
\end{equation}
and
\begin{equation}
V_{s}(x) =\frac{N_{s}(x)}{N_{s}(1)}V_{\rm mix}(1),
\end{equation}
where $V_{\rm mix}(x)$ is the volume of the mixture with mole fraction $x$.

We define the excess volume of the solution with respect to the ideal
mixture as 
\begin{equation}
\bigtriangleup V =\frac{V_{\rm mix}(x)}{V_{J}(x)+V_{s}(x)}-1
\end{equation}
If the excess volume $\bigtriangleup V$ is negative, the volume of the
solution is less than the volume of the ideal mixture. If it is
positive, the system expands after mixing at fixed temperature and pressure. In most contexts, we
use the volume fraction
\begin{equation}
\varphi = \frac{V_{s}(x)}{V_{J}(x)+V_{s}(x)}
\end{equation}
rather than the mole fraction $x$ to express different solute
concentrations of solutions.

We compare our simulation results with the data from experiments
\cite{LamaJCED1965,Mcglashan1976}, where the excess volume is expressed
in terms of $\bigtriangleup Y=\frac{V_{mix}-(n_{w}v_{w}+n_{m}v_{m})}{n_{w}+n_{m}}$, 
where $n_{m}$ is the number of the moles of methanol, $n_{w}$ is the
number of moles of water, the mole fraction is $x =
\frac{n_{m}}{n_{w}+n_{m}}$, and $v_{w}$ and $v_{m}$ are the molar volumes of water and methanol, respectively, at specific temperature
and pressure conditions. The conversion formulas between
$\bigtriangleup V$ and $\bigtriangleup Y$, and $\varphi$ and $x$ are
\begin{equation}
\bigtriangleup V =\frac{\bigtriangleup Y}{xv_{m} + (1-x)v_{w}}
\end{equation}
and
\begin{equation}
\varphi = \frac{x}{x+(1-x)\frac{v_{w}}{v_{m}}},
\end{equation}

Density is an important system property. 
For this purpose, we assume that the Jagla particles and
the solute particles correspond to the same number of water and methanol
molecules in a pure solution, respectively, and express the density of the
pure solute in terms of the density ratio
\begin{equation}
\rho = \frac {\frac{32}{v_{m}}}{\frac{18}{v_{w}}},
\end{equation}
where $v_{J}=\frac{V(0)}{N_{J}(0)}$ and $v_{s}=\frac{V(1)}{N_{s}(1)}$
are the volume per particle of the pure solvent and the pure solute,
respectively. We compare the simulation  with the experimental
number $\rho=0.79$.

The excess enthalpy is usually defined as
\begin{equation}
\bigtriangleup H_{e}=\frac{H_{\rm mix}-H_{m}-H_{w}}{n_{m}+n_{w}},
\end{equation}
where $H_{m}$ is
the total enthalpy of $n_{m}$ moles of pure methanol, and $H_{w}$ is the
total enthalpy of $n_{w}$ moles of pure water under specific temperature and
pressure conditions. To put the enthalpy comparison on the
same footing as the excess volume data, we choose to also report the
excess enthalpy on a volumetric basis and define the excess enthalpy per
volume as
\begin{equation}
\bigtriangleup H_{s}=\frac{H(x)-\frac{N_{s}(x)}{N_{s}(1)}H(1)-
  \frac{N_{J}(x)}{N_{J}(0)}H(0)}{V_{J}(x)+V_{s}(x)},
\end{equation}
where $H(x)$ is the enthalpy of the system with a mole fraction $x$.
The conversion formula is
\begin{equation}
\bigtriangleup H_{s}= \frac{\bigtriangleup H_{e}}{x(v_{m}-v_{w})+v_{w}}.
\end{equation}

In our simulation, we measure $\bigtriangleup H_{s}$ in units of
$U_{o}/a^3$.  In order to compare our results with experimental data, we
need to convert our units into J/cm$^3=$~MPa. In accordance with
Ref.~\cite{YanPRE2008}, we use $U_{o}=4.75$KJ/mol and $a=2.7\times
10^{-8}$cm.  Then we convert by simply multiplying our simulation
results by $4.008\times 10^2$.

\section{RESULTS and DISCUSSION}
\subsection{Effects of the parameters on model behavior}

\begin{figure}[H]
\centerline{\includegraphics[width=0.8\textwidth]{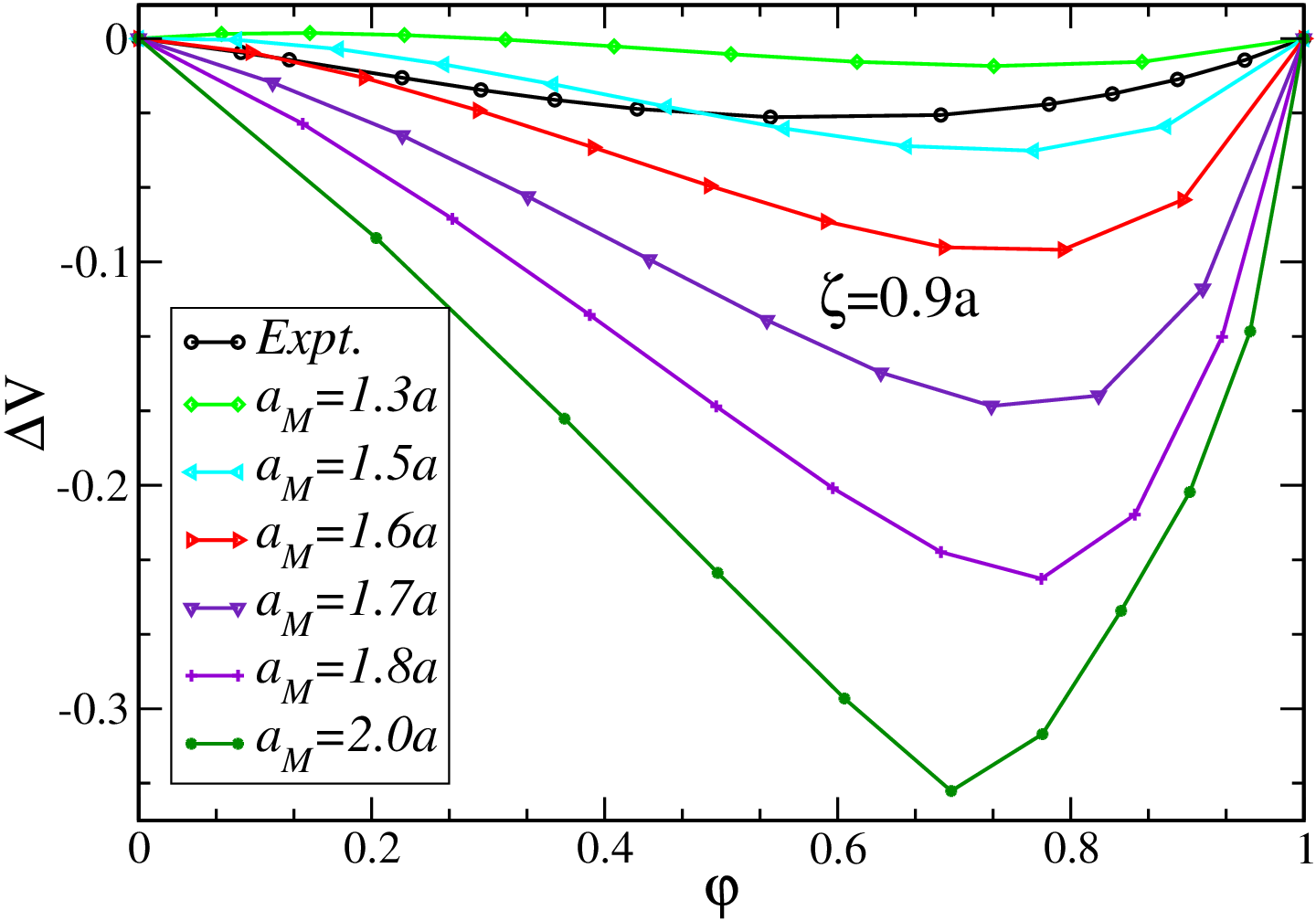}}
\caption{(color online) Dependence of the excess volume as a function of
  the solute volume fraction for a range of hard core diameters
  $a_{\rm M}$, for the dimer model with $\zeta=0.9a$. The simulations are
  done at $T=0.5$ and $P=0.02$. The experimental result (circles) are for
  $T=298$K and $P=0.1$MPa \cite{Mcglashan1976}, where the maximum volume fraction
  reduction occurs at $\varphi_{\rm max}\approx 59.7\%$.  For the simulation results,
  when $a_{\rm M}\leq 1.3a$, the excess volume has both positive and negative values. As $a_{\rm M}$ increases, the
  excess volume becomes more negative and $\varphi_{\rm max}$, the maximum
  reduction volume fraction, decreases.}
\label{dimerhardsphere}
\end{figure}

Because there are several parameters in our dimer model, we first
investigate how these affect the simulations, searching for a
set of parameters that can best model ambient methanol. Since the goal
is to explore the excess properties of the model, vis a vis real
methanol-water solution, we compare our simulation results with the
results reported in Ref.~\cite{Mcglashan1976} concerning the excess
volume at $T=298$K and $P=0.1$~MPa.  We set our simulation temperature
and pressure at $T=0.5$ and $P=0.02$, and we change the diameter of the
hard spheres that model the methyl group in methanol.  In
Fig.~\ref{dimerhardsphere}, we show the excess volume of the
solutions with different hard core diameters of the hard spheres $a_{\rm
  M}$, for a fixed bond length $\zeta=0.9a$, across the entire range of amphiphilic
solute volume fraction. If $a_{\rm M}\leq 1.3a$, we cannot reproduce the
volume reduction over the entire range of volume fraction. If $a_{\rm
  M}>1.3a$, the mixture shrinks over the entire range of volume
fractions. When the hard core diameter increases, the volume fraction corresponding to 
maximum volume reduction (henceforth referred to as maximum reduction volume fraction) decreases. 
However, for this value of $\zeta$ we cannot reproduce the experimentally-observed maximum
reduction volume fraction.

\begin{figure}[H]
\includegraphics[width=0.5\textwidth]{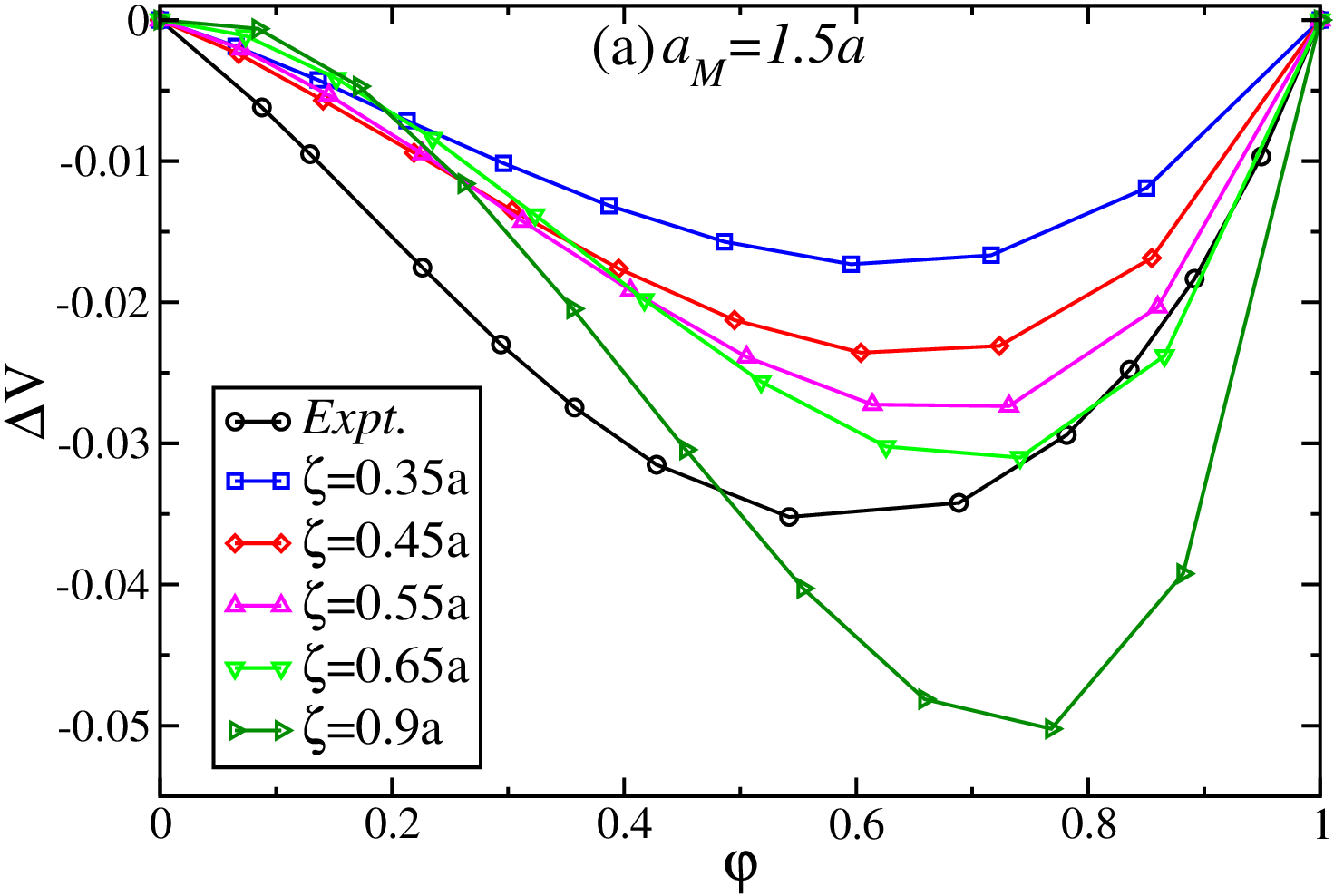} \\
\includegraphics[width=0.5\textwidth]{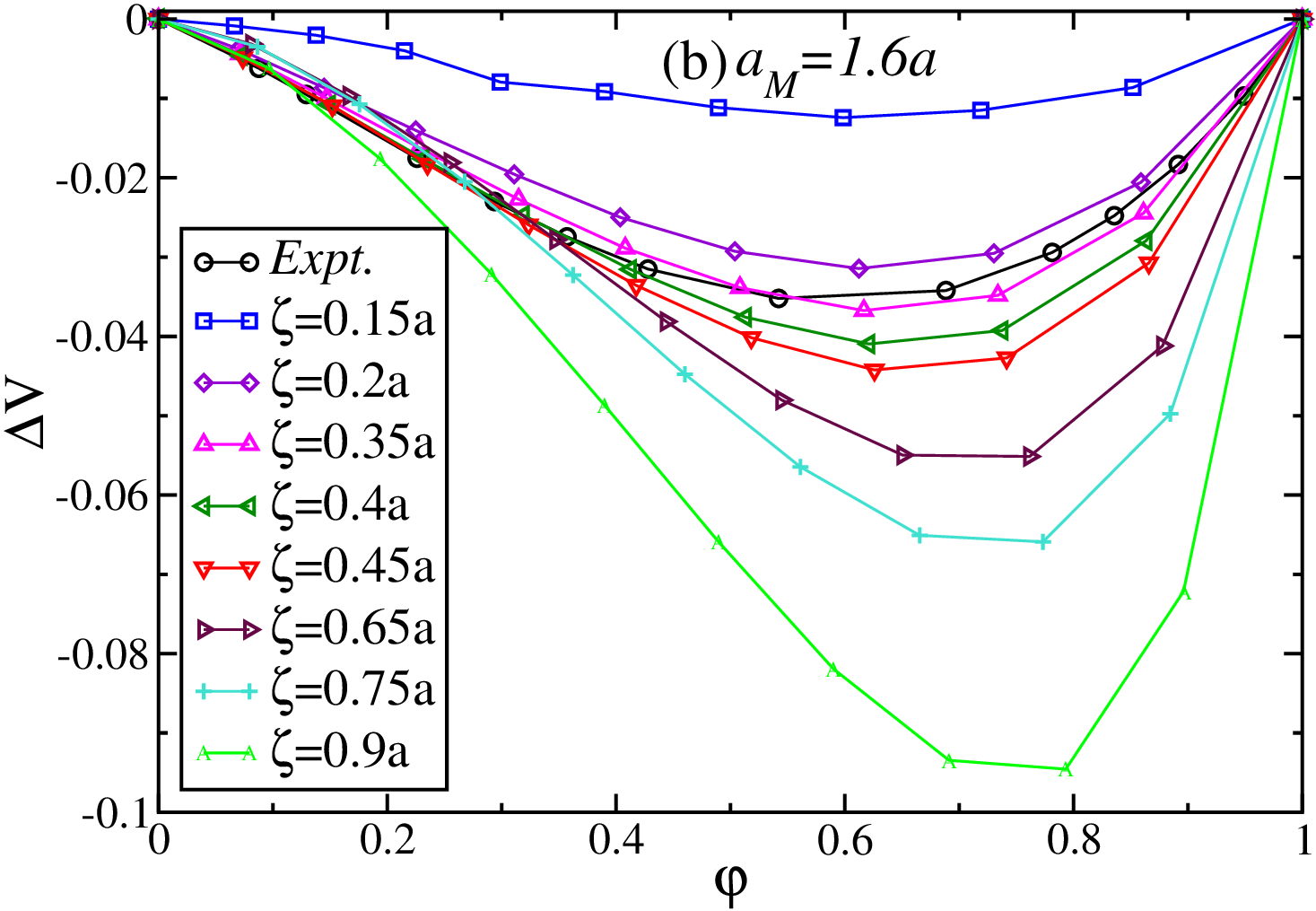}  \\
\includegraphics[width=0.5\textwidth]{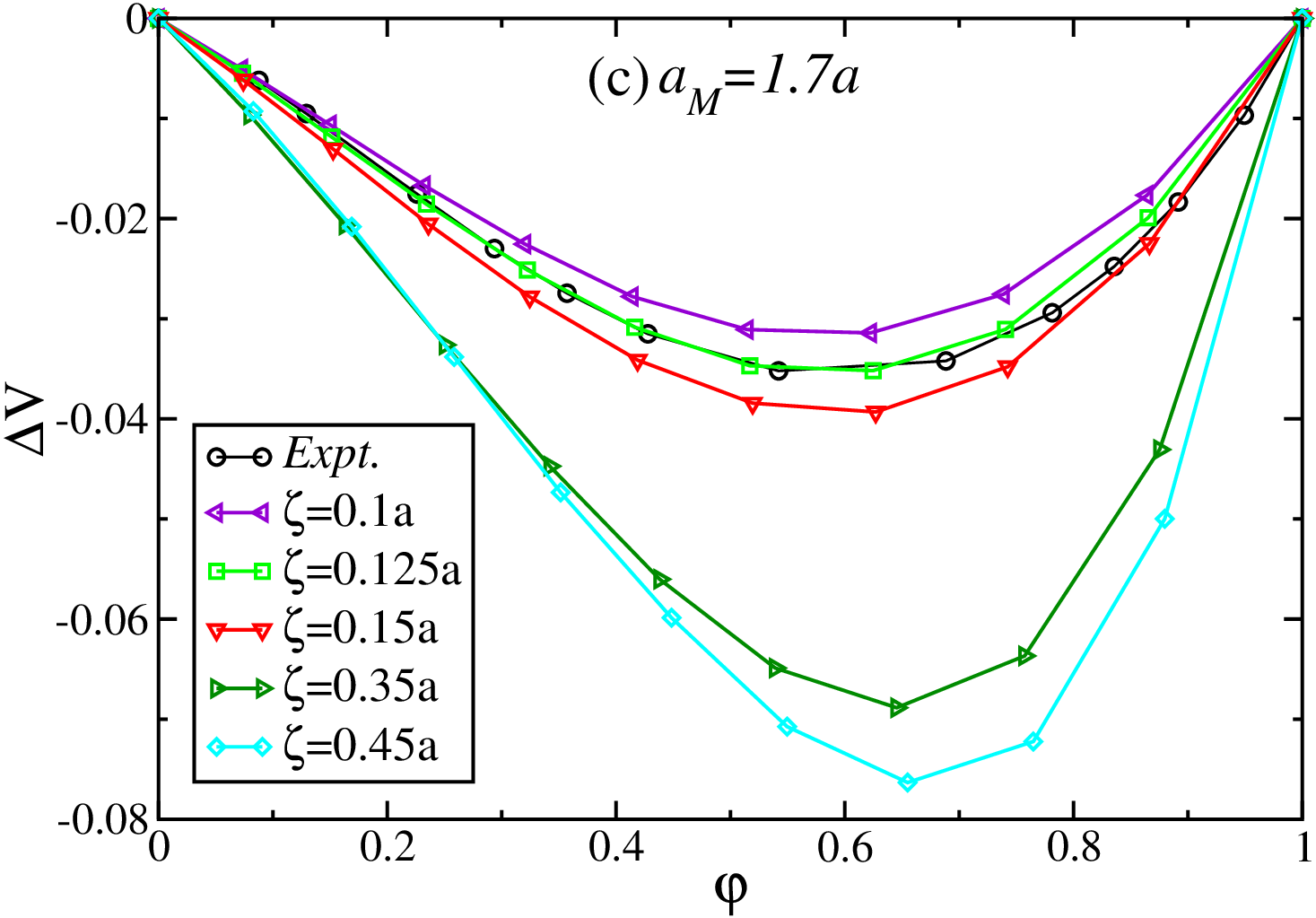}
\caption{(color online) Dependence of the excess volume on solute
  volume fraction for various solute bond lengths $\zeta$ at three
  values of hard core diameter $a_{\rm M}$. The simulations are done at
  $T=0.5$ and $P=0.02$. (a) $a_{\rm M}=1.5a$, (b) $a_{\rm M}=1.6a$, (c)
  $a_{\rm M}=1.7a$.  As $\zeta$ increases, the excess volume,
  $\bigtriangleup V$ becomes more negative and the maximum reduction
  volume fraction $\varphi_{\rm max}$ increases.  Calculations performed 
  using the parameter sets $\zeta=0.35a$, $a_{\rm
    M}=1.6a$ in (b) and $\zeta=0.125a$, $a_{\rm M}=1.7a$ in (c) agree
  well with experiment.}
\label{dimerbond}
\end{figure}

We next show, in Fig.~\ref{dimerbond}, how bond length affects 
the excess volume.  We choose three different hard sphere diameters
and vary the bond length. For $a_{\rm M}=1.5a$ [Fig.~\ref{dimerbond}(a)],
as the bond length increases, the excess volume becomes more negative
and the maximum reduction volume fraction increases, but its value is
always larger than in experiments. For $a_{\rm M}=1.6a$
[Fig.~\ref{dimerbond}(b)] and $a_{\rm M}=1.7a$
[Fig.~\ref{dimerbond}(c)], the excess volume and the maximum reduction
volume fraction exhibit the same trend as the bond changes as seen for 
$a_{\rm M}=1.5a$. However, their maximum reduction volume fraction is
smaller and closer to the experimental data.  Thus, we can see
that the maximum reduction volume fraction $\varphi_{\rm max}$ and the
corresponding $\bigtriangleup V_{\rm max}$ agree well
with the experimental observations for selected bond lengths; i.e., for $a_{\rm M}=1.6a$ and
$\zeta=0.35a$, $\varphi_{\rm max}=63.6\%, \bigtriangleup V_{\rm max}= -3.67\%$, 
and for $a_{\rm M}=1.7a$ and $\zeta=0.125a$, $\varphi_{\rm max}=58.717\%$,
$\bigtriangleup V_{\rm max}= -3.55\%$. In addition,
over the entire range of volume fractions the excess volume agrees quantitatively with
experiment for these two sets of parameters.

\begin{figure}[H]
\centerline{\includegraphics[width=0.8\textwidth]{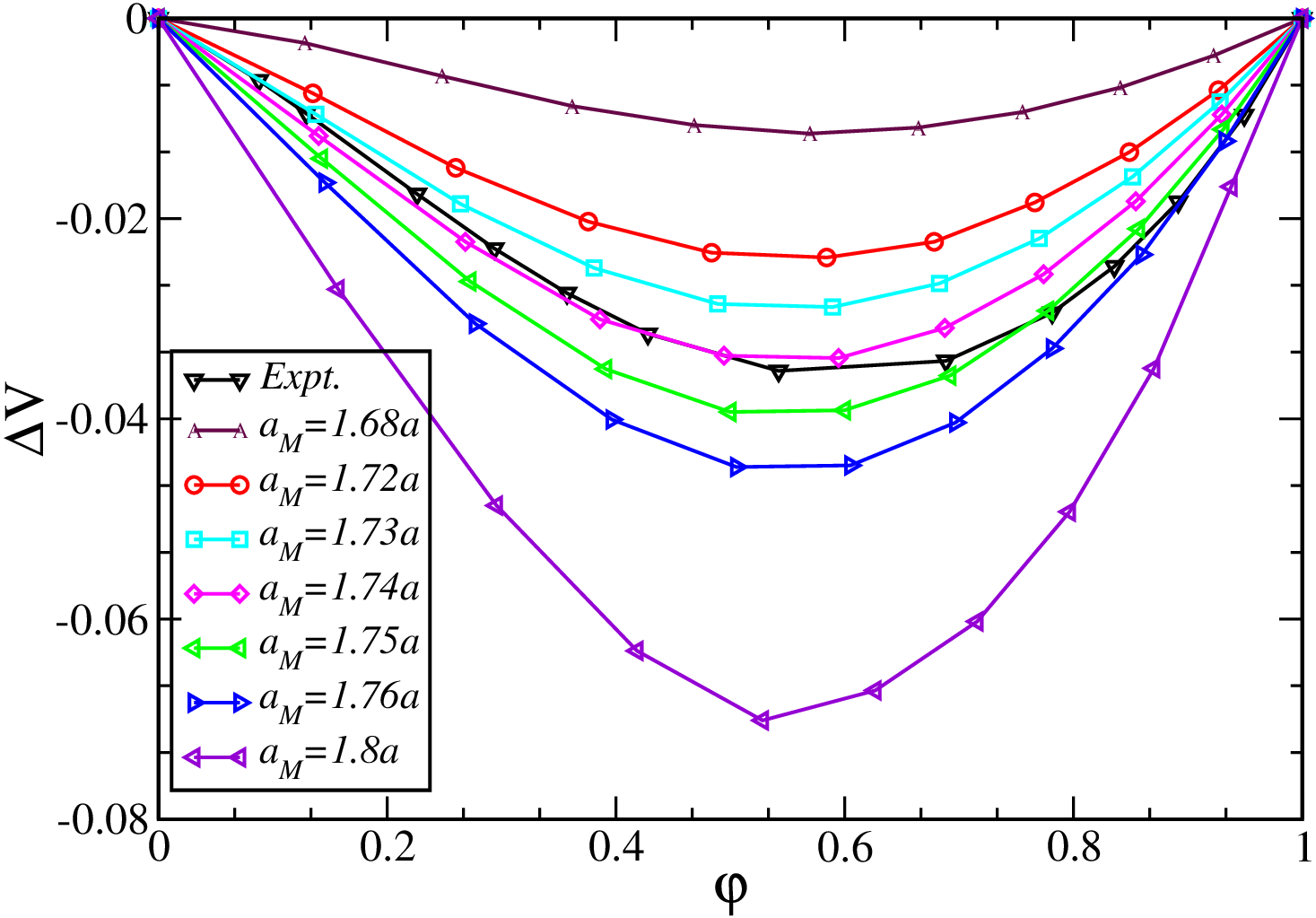}}
\caption{(color online) Dependence of the excess volume on solute
  volume fraction for various hard core diameter values ($a_{\rm M}$), in the
  monomer ($\zeta=0$) model. Simulations at $T=0.5$ and $P=0.02$. As
  $a_{\rm M}$ increases, $\bigtriangleup V$ becomes more negative; the
  maximum reduction volume fraction $\varphi_{\rm max}$ is relatively
  insensitive to $a_{M}$. The excess volume at $a_{\rm M}=1.74a$ is
  similar to experiment, but the results are  less accurate than for the dimer model [see Fig.~\ref{dimerbond}(b) and \ref{dimerbond}(c)].}
\label{volumemonomer}
\end{figure}

In the dimer model, we achieve the closest agreement with the
experimental data at $a_{\rm M}=1.7a$ and $\zeta=0.125a$.  At this set
of parameters, the hard sphere and the Jagla particle nearly overlap.
The dimer model can thus be modeled as a monomer with a hard core $a_{\rm
  M}\approx 1.7a$, as shown in Eqs.~(5) and (6). In order to achieve the
closest agreement with the experimental data, we vary the single
parameter, $a_{\rm M}$ in this model (Fig.~\ref{volumemonomer}).  For
all cases of $a_{\rm M}$, the reduction property can be reproduced across
the entire range of volume fractions. At $a_{\rm M}=1.73a$ and $a_{\rm
  M}=1.75a$, the trend of the excess volume is very similar to the
experimental results.  For $a_{\rm M}=1.74a$, the maximum reduction
volume fraction is close to the experimental data, and in the range of
volume fractions smaller than the maximum reduction volume fraction, the
excess volume is also quantitatively reproduced. Overall, the monomer
model can reproduce the excess volume curve qualitatively but is not as
accurate as the dimer model.

At $T=298$K and $P=0.1$~MPa, the methanol density is 0.78663 g/cc
\cite{Mcglashan1976}, and its ratio with respect to water is 0.79. We
have explored the relative density of the neat amphiphilic solutes with
respect to a pure Jagla solvent when we change the parameters. The
density increases when we shorten the bond length, when the diameter of
the hard sphere $a_{\rm M}$ in the dimer model decreases, and when the
hard core of the monomer $a_{\rm M}$ decreases.  From the above-discussed results
we know that when $a_{\rm M}=1.7a$ and $\zeta=0.125a$, we achieve the
best agreement between our simulation results and the experimental
results. However, the density ratio of the pure solute for this set of
parameters is 0.66, less than the density ratio of real methanol at
$T=298$~K and $P=0.1$~MPa. In the monomer model, when the hard core
diameter of monomer is $a_{\rm M}=1.74a$, the density ratio is 0.64,
also less than the experimental value.

\subsection{Temperature and pressure dependence of the excess volume}

\begin{figure}[H]
\includegraphics[width=0.5\textwidth]{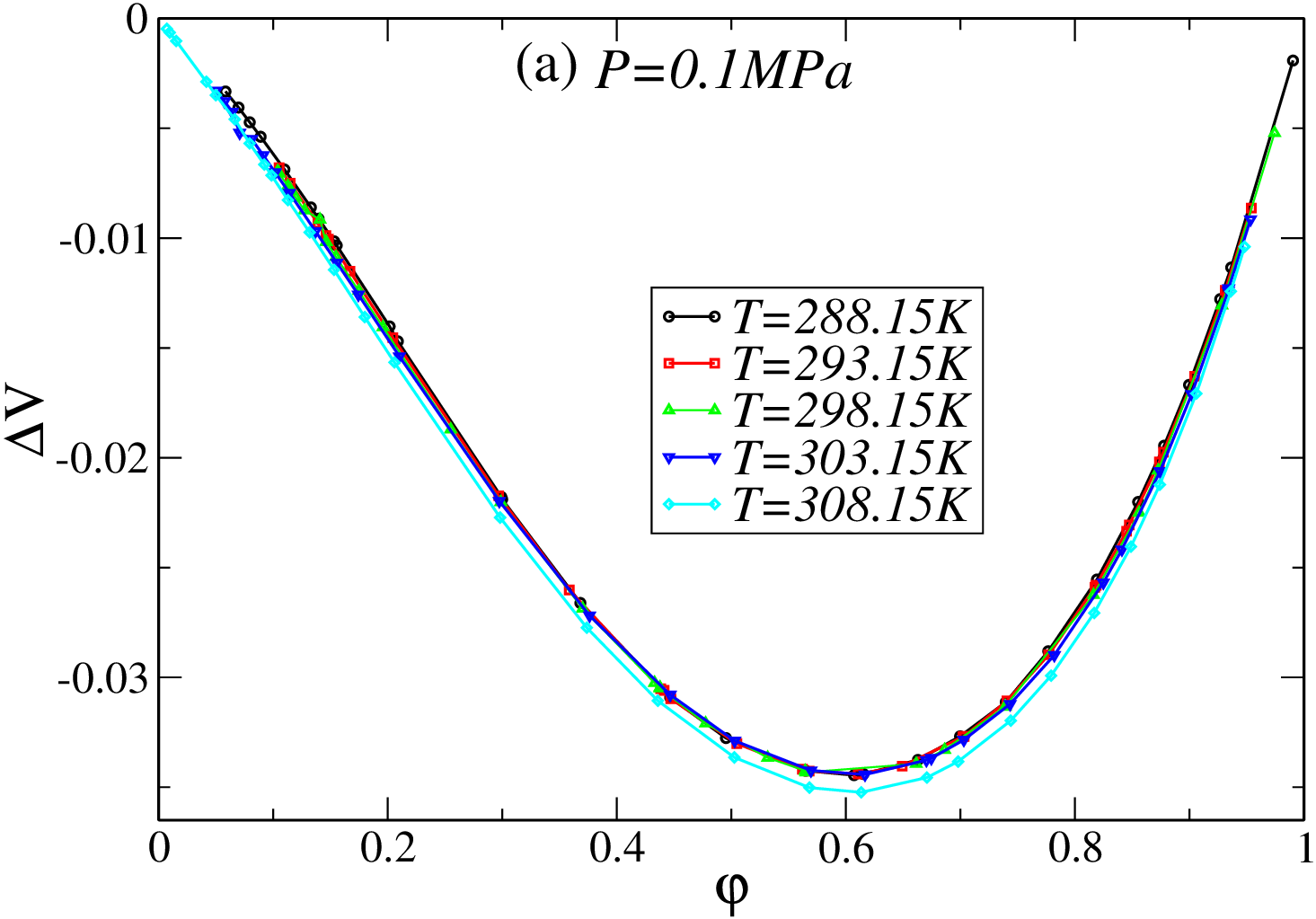} \\
\includegraphics[width=0.5\textwidth]{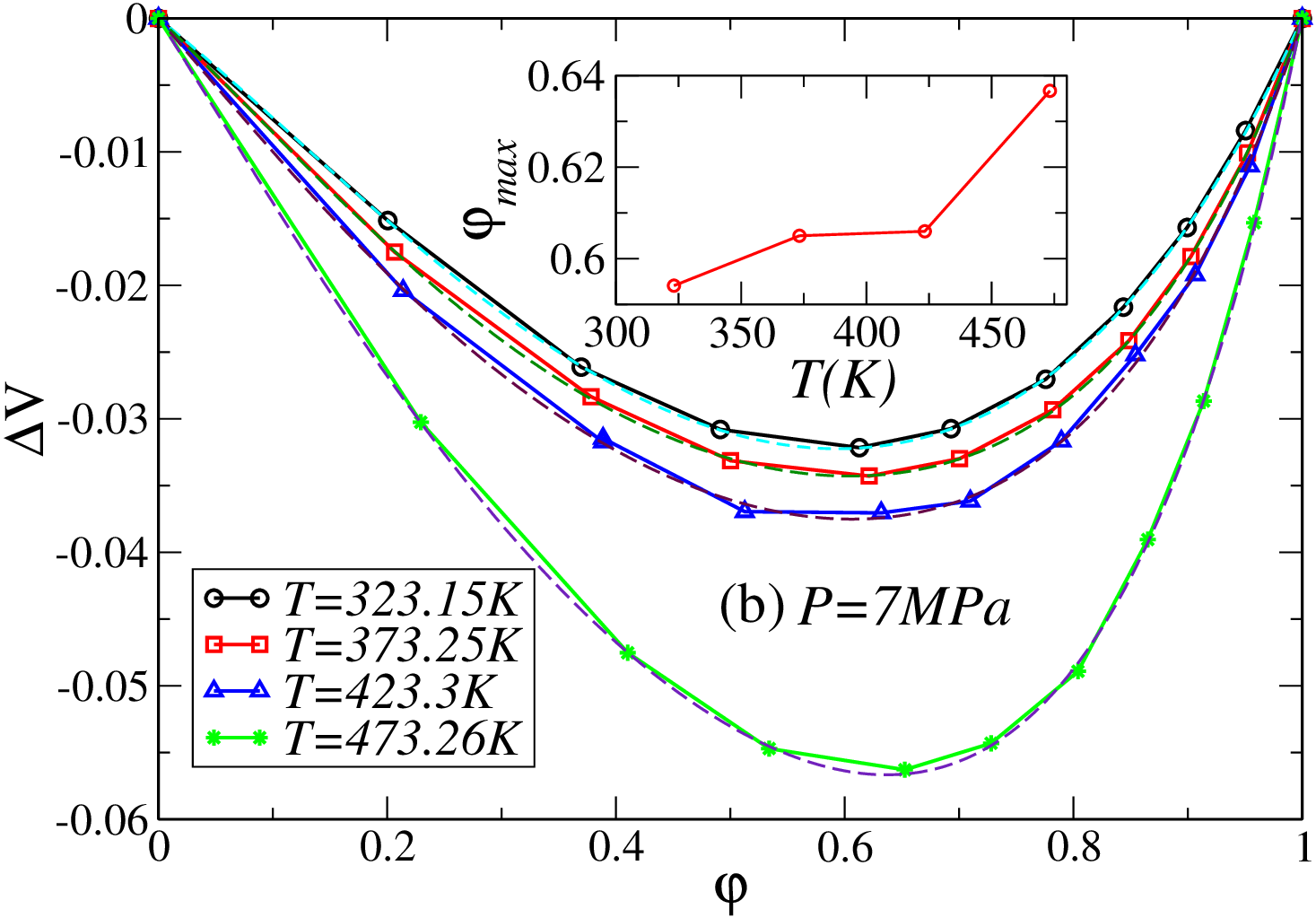} \\
\includegraphics[width=0.5\textwidth]{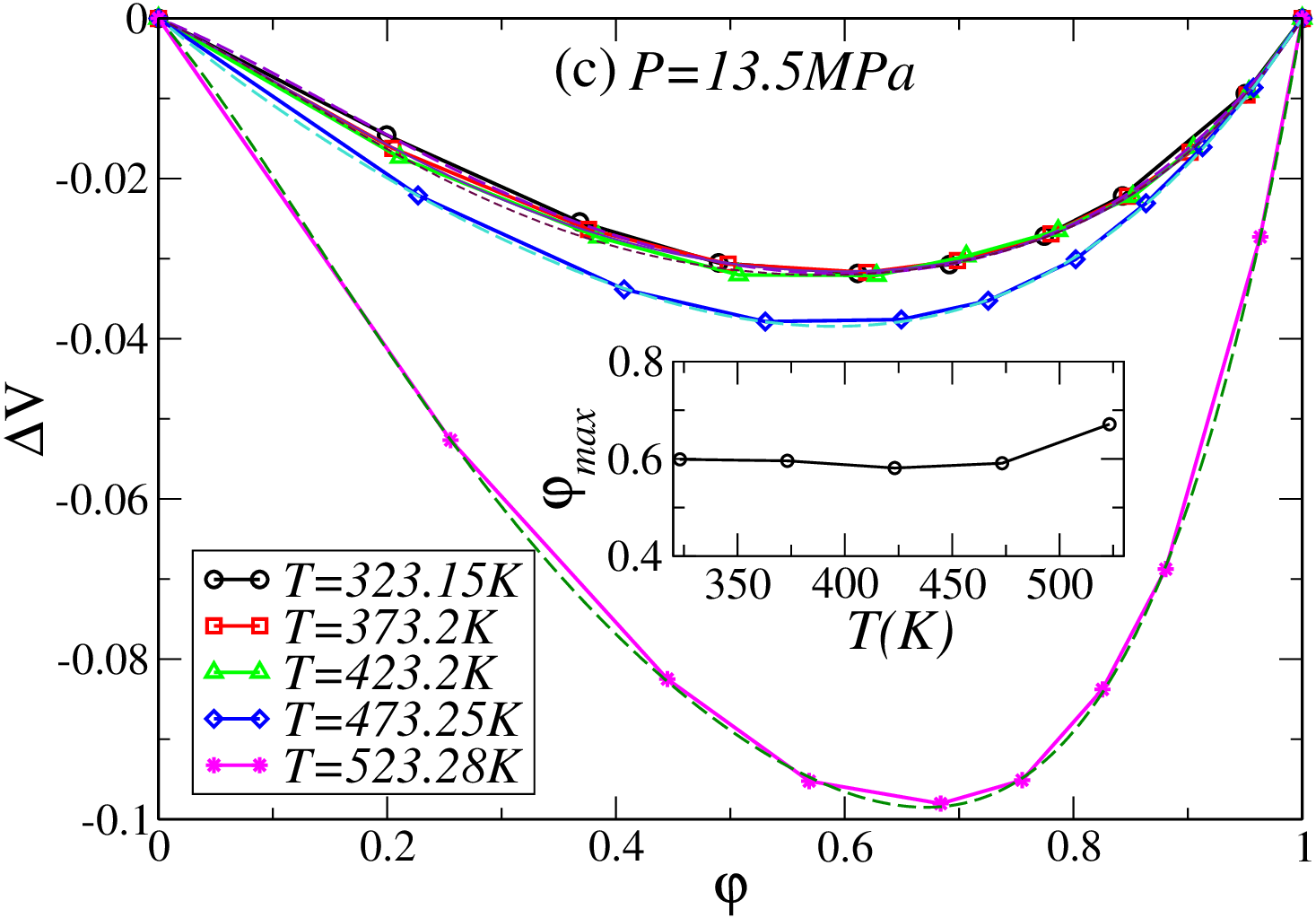}
\caption{(color online) Experimental data \cite{BensonJSC1980,XiaoJCT1997} on the temperature dependence of the excess volume,
  as a function of the volume fraction of methanol, $\varphi$, at different
  pressures (Dotted lines are polynomial fits of experimental data): (a)
  $P=0.1$MPa, (b) $P=7$MPa, and (c) $13.5$MPa. Note that in (a)
  temperature interval between two measurements is 5K, while the
  temperature difference in (b) and (c) is 50K. The excess volume is the relative difference between the volume occupied by the mixture and the sum
of the volumes of the pure components before mixing, at fixed temperature and pressure}
\label{temperatureexperiment}
\end{figure}

Experimentally, when the temperature change is small, e.g.,
5K, the change of the maximum reduction volume fraction is negligible and
there is a small increase in $\bigtriangleup V$ as the temperature
increases [Fig.~\ref{temperatureexperiment}(a)].  Over large temperature
intervals, e.g., 50K, the excess volume becomes increasingly negative
and the maximum reduction volume fraction increases slightly as the
temperature increases [Figs.~\ref{temperatureexperiment}(b) and
  \ref{temperatureexperiment}(c)].  We use our two models to check the
temperature and pressure dependence of the excess volume. Choosing the
parameters that can best reproduce the experimental results of the
excess volume at $T=298$~K and $P=0.1$~MPa, we set $a_{\rm M}=1.7a$ and
$\zeta=0.125a$ for the dimer model.

\begin{figure}[H]
\includegraphics[width=0.5\textwidth,angle=0]{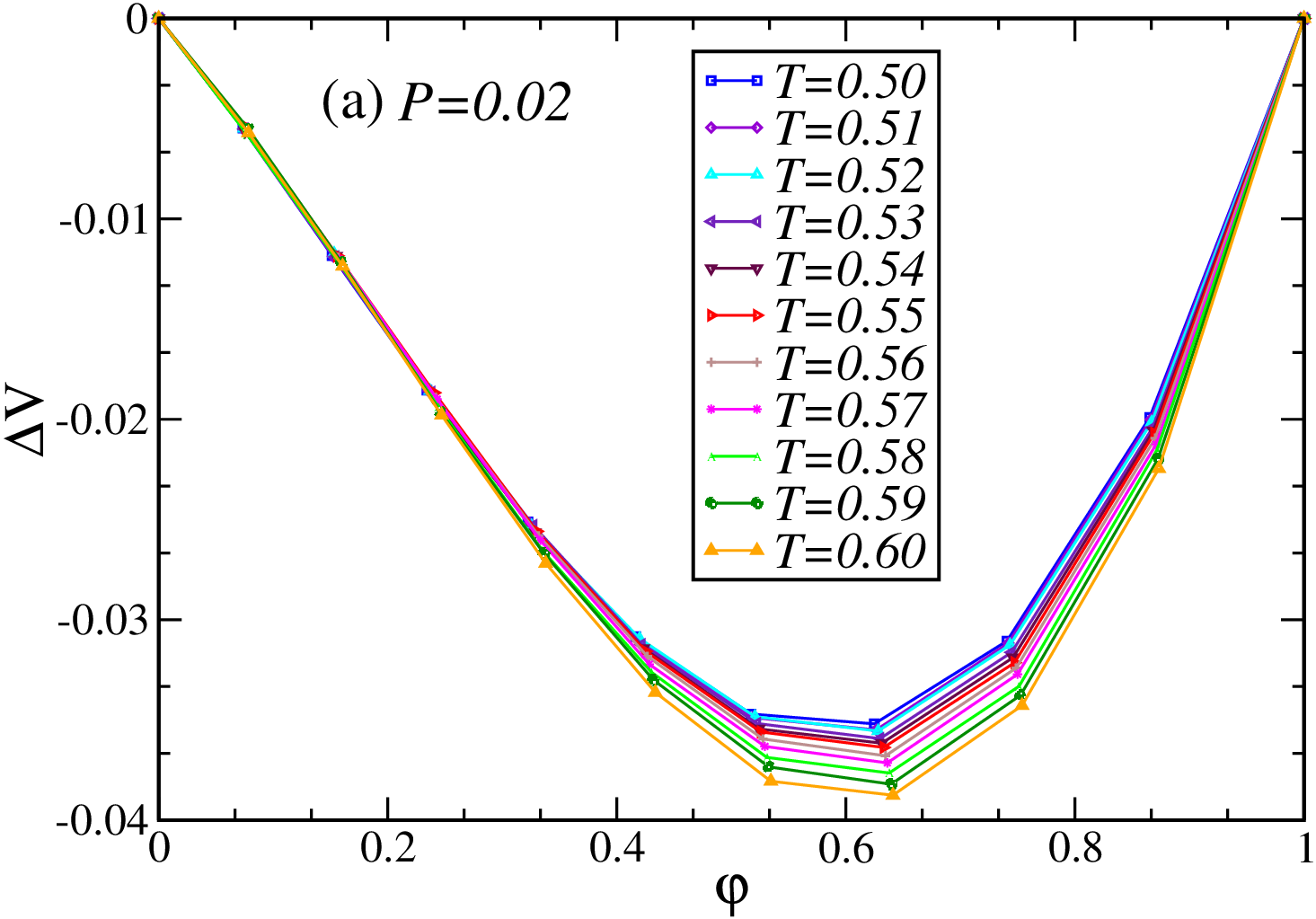}\\ 
\includegraphics[width=0.5\textwidth,angle=0]{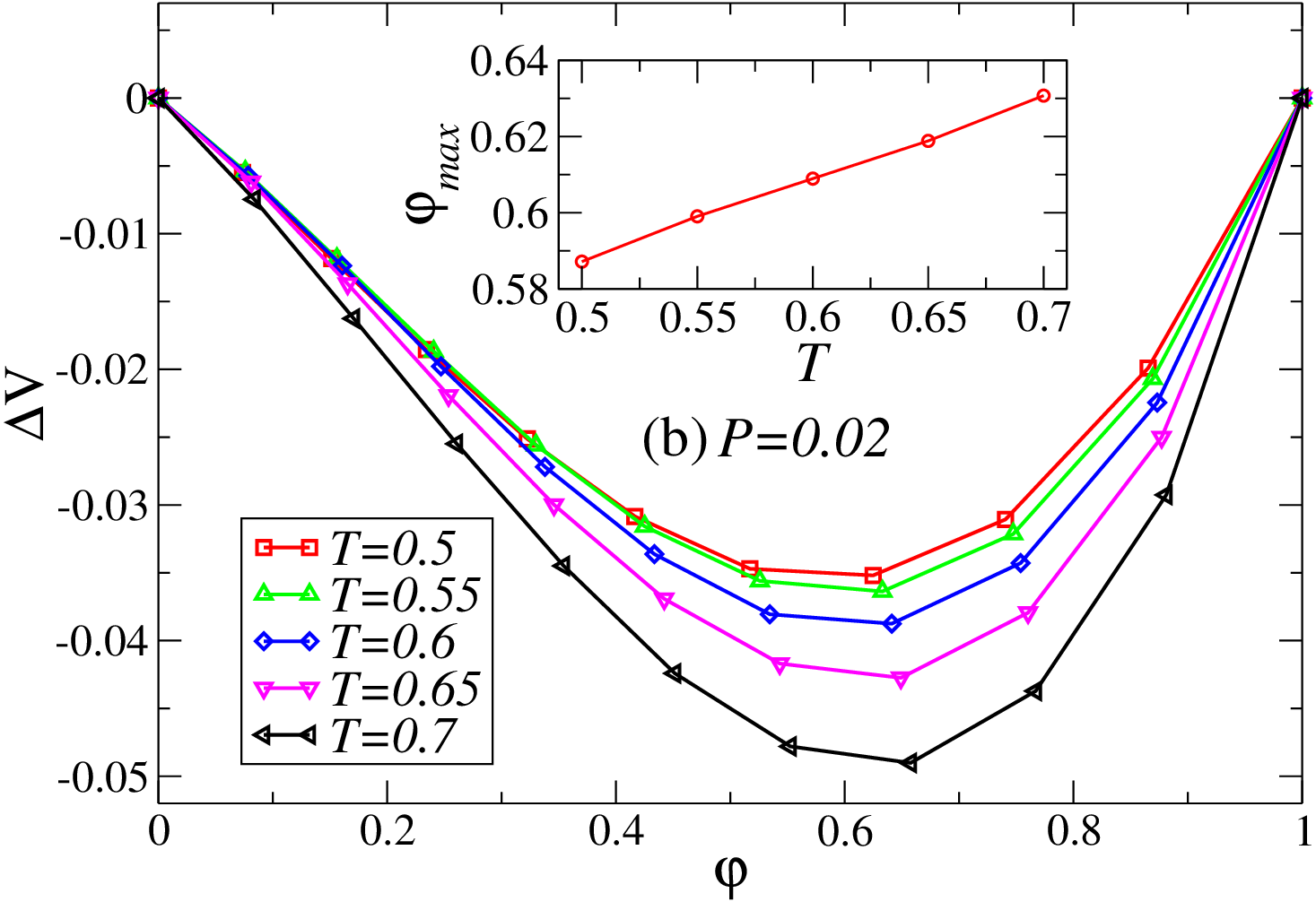}\\ 
\includegraphics[width=0.5\textwidth,angle=0]{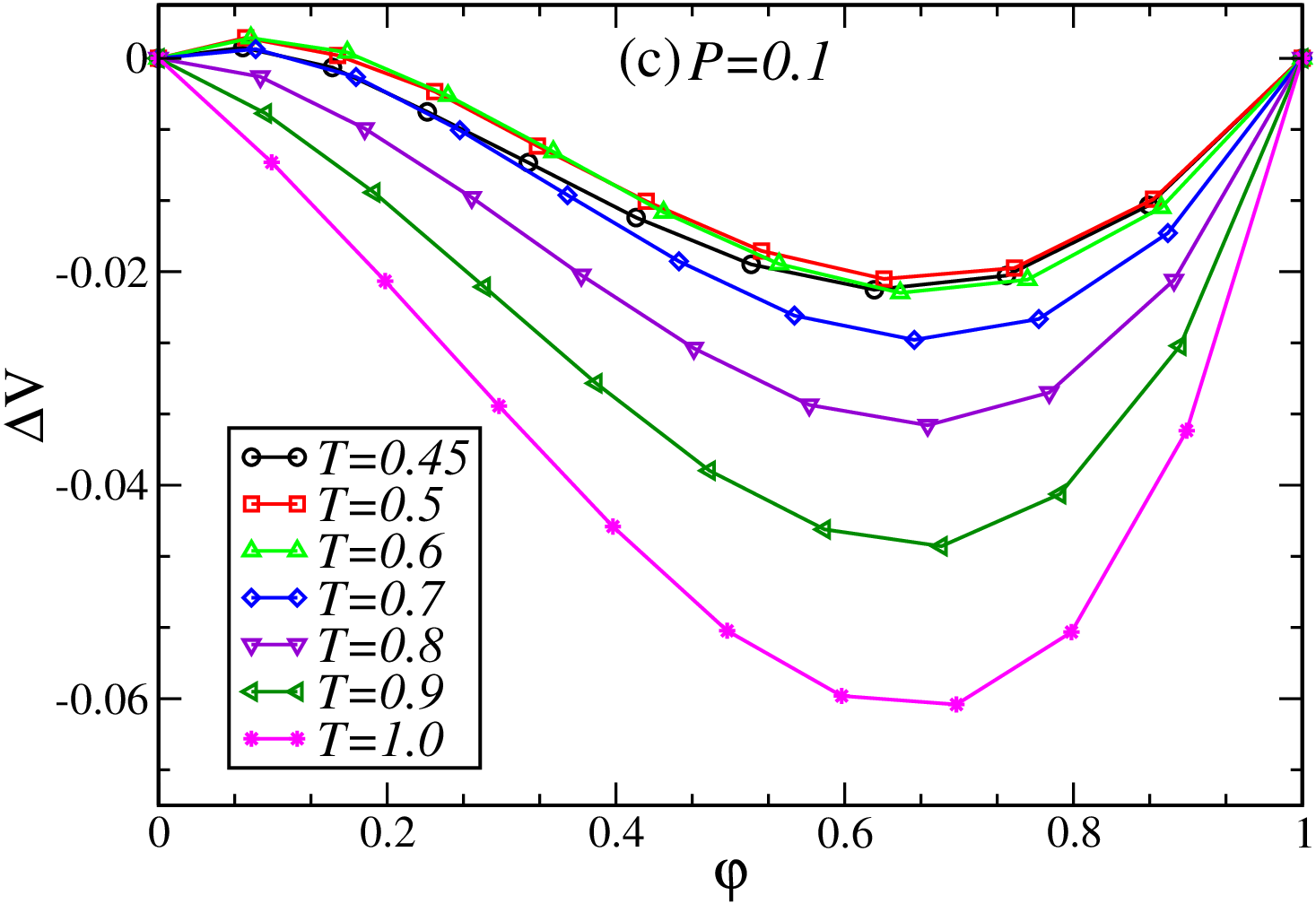}
\caption{(color online) Dependence of the excess volume on solute volume fraction, 
  at various temperatures and three different pressures, for the dimer model with $a_{\rm M}=1.7a$, $\zeta=0.125a$.  (a)
  $P=0.02$, which is comparable to atmospheric pressure. The temperature difference between curves
  is 0.01.  (b) $P=0.02$. The temperature difference between
  curves is 0.5.  (c) $P=0.1$.  
  As the temperature increases, the excess
  volume becomes more negative and the maximum reduction volume fraction
  increases. In (c), when the temperature is low, there is a range of dilute mixtures for which $\bigtriangleup V > 0$. 
  There are no
  corresponding experimental observations.  }
\label{temperaturedimer}
\end{figure}

We calculate the temperature dependence of the excess volume at
pressures $P=0.02$ and $P=0.1$ in the dimer model
[Fig.~\ref{temperaturedimer}]. At $P=0.02$ ($\approx 1.0$~atm), we first
measure the excess volume at a series of temperatures separated by 0.01
in the range of $T=0.5$--0.6 [Fig.~\ref{temperaturedimer}(a)]. The
excess volume becomes slightly more negative and the change of the
maximum reduction volume fraction is negligible as the temperature
increases.  When we enlarge the temperature interval to 0.05
[Fig.~\ref{temperaturedimer}(b)], the excess volume becomes increasingly
negative and the maximum reduction volume fraction increases noticeably
as the temperature increases. At $P=0.1$
[Fig.~\ref{temperaturedimer}(c)], we find approximately the same
results. However, when the temperature is approximately 0.5, the excess
volume of the solution with less than 15 percent of amphiphilic solutes
goes to zero or becomes positive, indicating that in our model the
volume does not decrease, and can even
expand at these conditions. We will discuss this ``bump'' in the excess volume curve later.
The temperature dependence at pressures $P=0.02$ and $P=0.1$ using the
$a_{\rm M}=1.74a$ monomer model (not shown) are comparable to those of
the dimer model.

\begin{figure}[H]
\includegraphics[width=0.7\textwidth,angle=0]{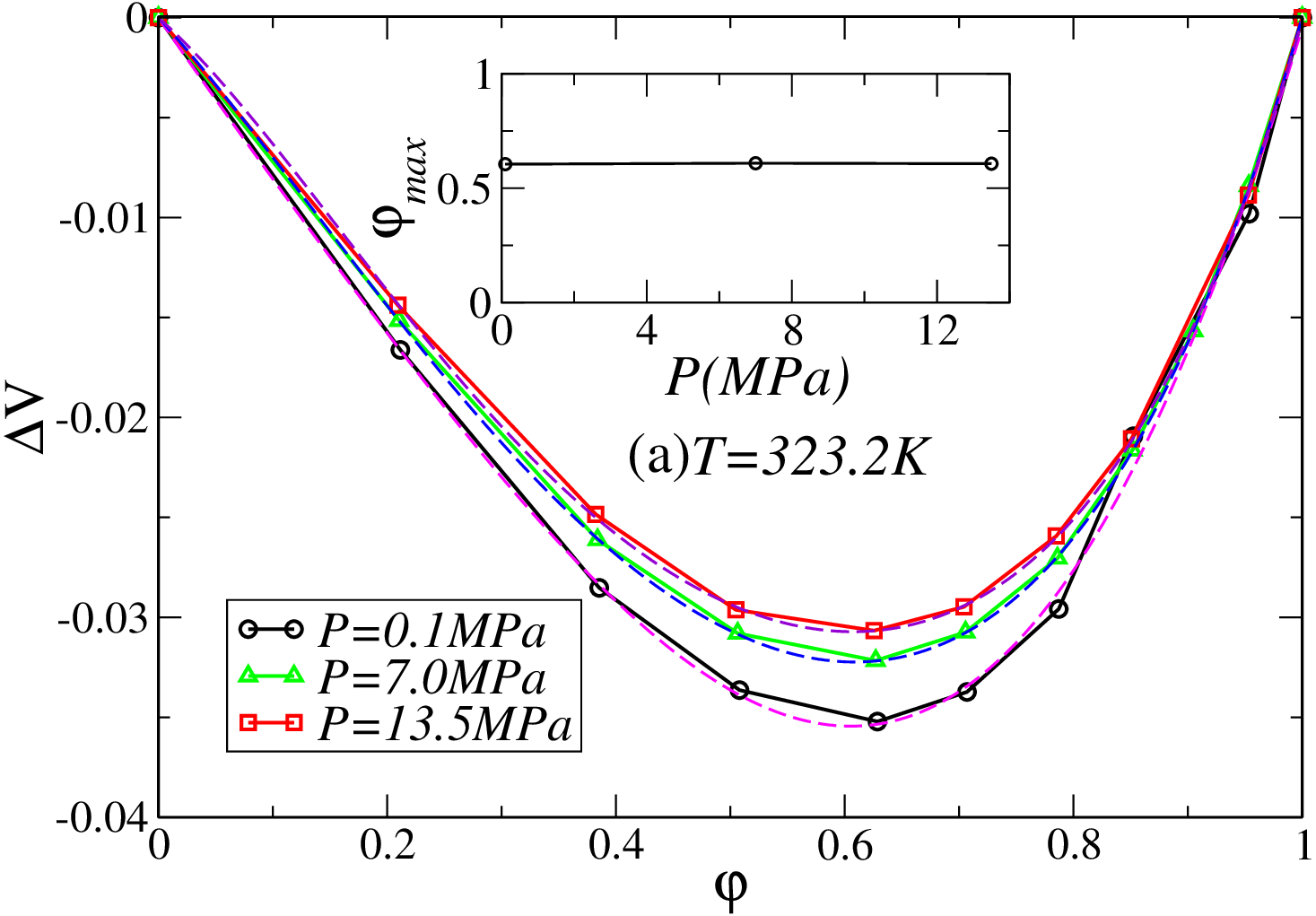} \\
\includegraphics[width=0.7\textwidth,angle=0]{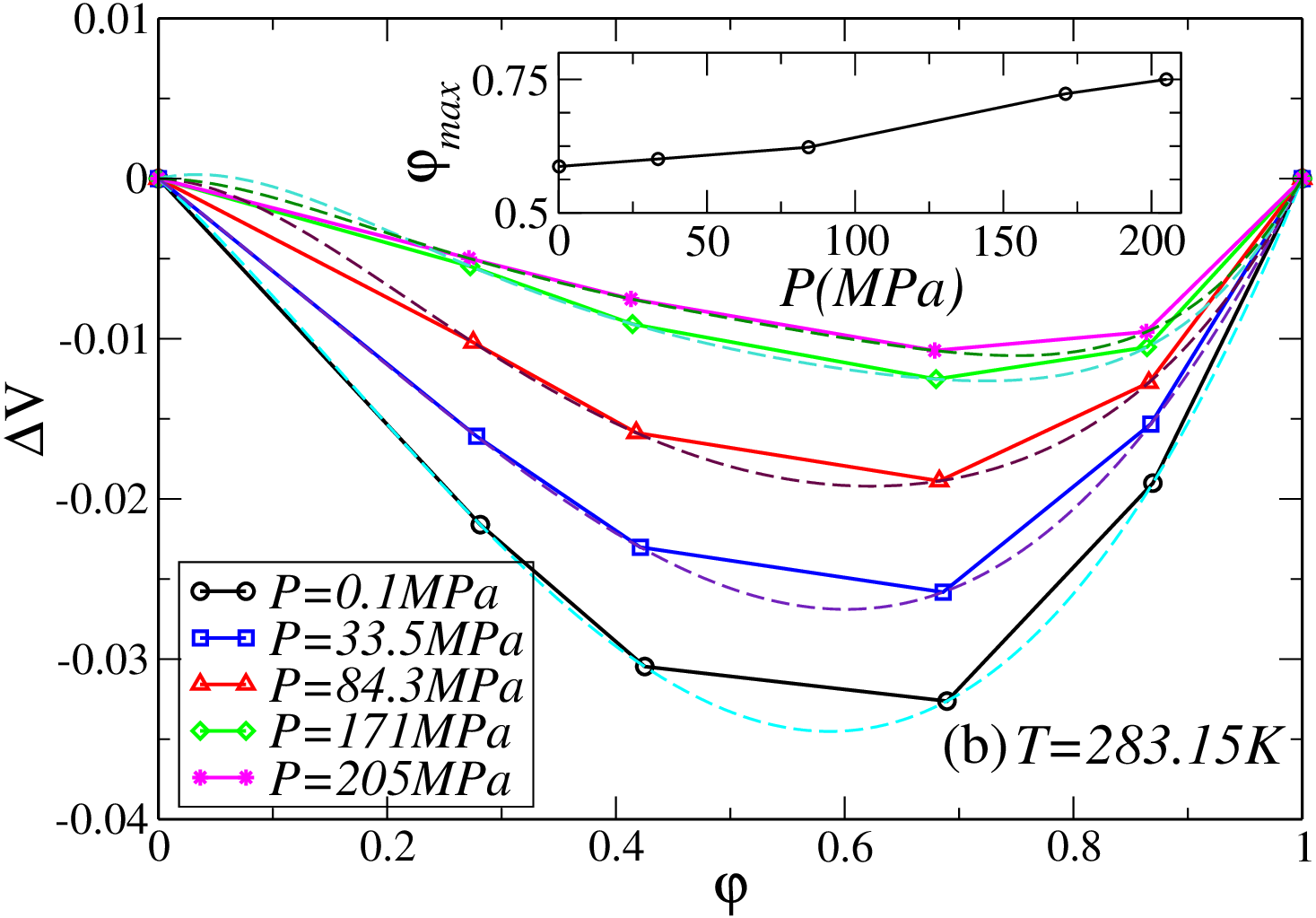}
\caption{(color online) Experimental data \cite{XiaoJCT1997,KubotaIJT1987} on the pressure dependence of the excess volume as a
  function of the volume fraction of methanol, $\varphi$, at two different
  temperatures (dotted lines are polynomial fits of experimental data): (a)$T=323.2$K and (b) $T=285.15$K. Note that the range of the pressure
  shown in (a) is smaller than that in (b). The excess volume is defined in the caption of Fig.~\ref{temperatureexperiment}}
\label{pressureexperiment}
\end{figure}

We next compare the calculated pressure dependence of the
excess volume with the experimental results. In
Fig.~\ref{pressureexperiment}, we report the experimental pressure
dependence of the excess volume at two different temperatures,
$T=323.15$K and $T=283.15$K, using two different pressure intervals.  The
results at $T=323.15$K cover pressures from $P=0.1$MPa to $P=13.5$MPa,
and those at $T=283.15$K cover pressures from $P=0.1$MPa to $P=205$MPa.
We can see that, as the pressure increases, the excess volume becomes
less negative. Regarding the maximum reduction volume fraction of the
excess volume, in Fig.~\ref{pressureexperiment}(a), its value does not
noticeably change with temperature, however in
Fig.~\ref{pressureexperiment}(b), we can clearly see that as the
pressure increases, the maximum reduction volume fraction does increase.

\begin{figure}[H]
\centerline{\includegraphics[width=0.8\textwidth,angle=0]{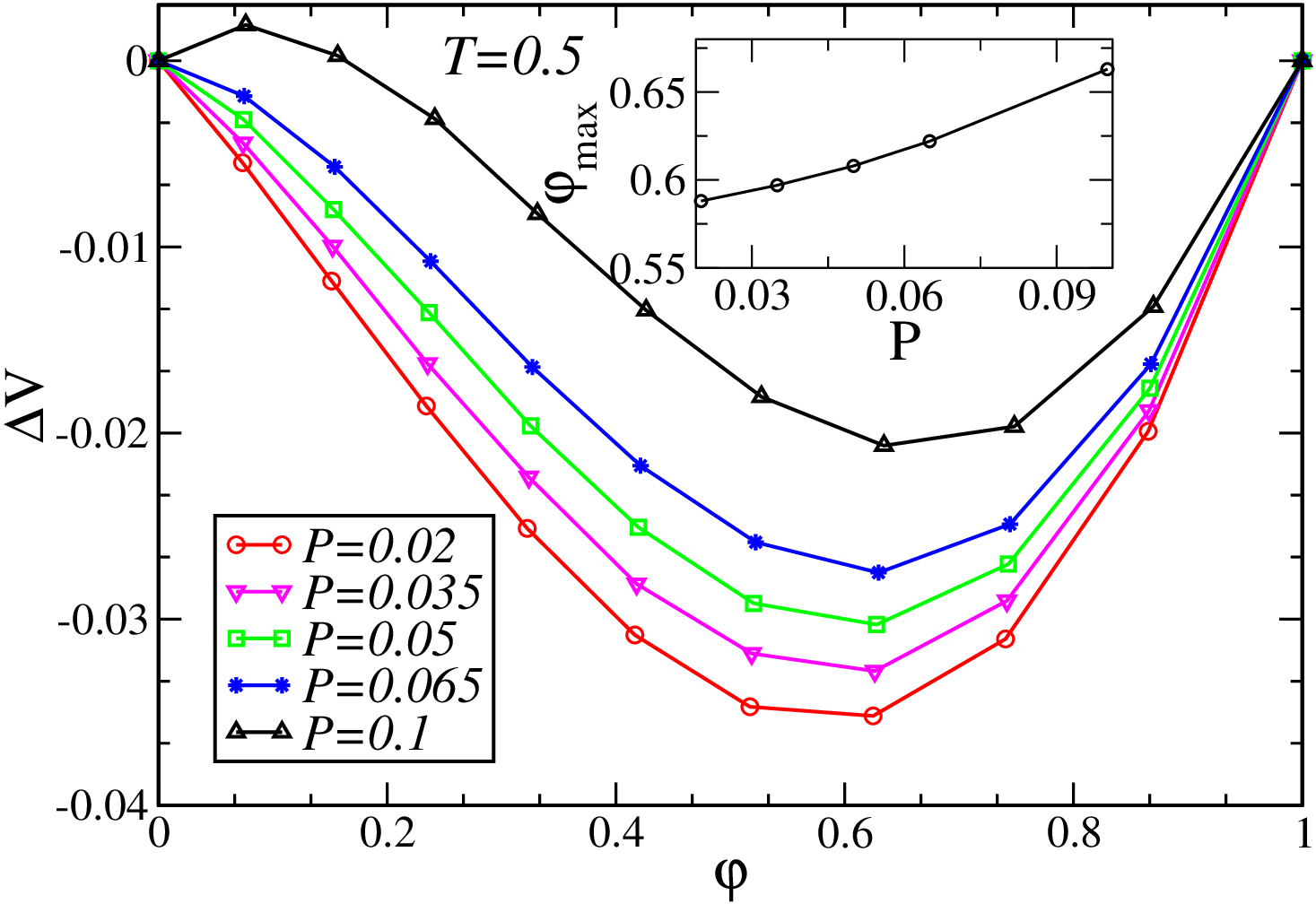}}
\caption{(color online) Dependence of the excess volume on solute volume fraction for several different
  pressures, at $T=0.5$, for the dimer model ($a_{\rm M}=1.7a$,
  $\zeta=0.125a$).  As the pressure increases, the excess volume becomes
  more negative and the maximum reduction volume fraction increases very
  slightly.}
\label{pressuredimerT05}
\end{figure}

In our simulation, we fix the temperature at $T=0.5$ and calculate the
excess volume at pressures from $P=0.02$ to $P=0.1$.  Our simulation
results for the dimer model (see Fig.~\ref{pressuredimerT05}) agree with
the experimental results quite well, namely, as the pressure increases
the excess volume becomes less negative and the maximum reduction volume
fraction increases. Regarding the positive ``bump'' at the small volume
fraction of amphiphilic solutes, we are not aware of 
experimental data for this range of volume fractions, so this result is
a prediction, namely that at high pressures and low temperatures, dilute
methanol-water solutions have a positive excess
volume.  The simulation results for the best monomer model are quite
similar to those for the dimer model except that the positive ``bump''
is smaller.

\subsection{Excess enthalpy}

The excess enthalpy in methanol-water solutions has been measured at
$T=298$~K, and $P=0.1$~MPa \cite{LamaJCED1965}. The
excess enthalpy is negative (exothermic mixing), consistent with the picture
of strong association that follows from the volumetric behavior. 
The maximum reduction occurs at a volume
fraction $\varphi_{\rm max}=42\%$, differing from that of the excess
volume at these specific conditions.  At pressures $P=0.1$MPa,
$P=20$MPa, and $P=39$MPa, the excess enthalpy becomes less negative as the temperature increases.
The maximum reduction volume fraction increases as the temperature
increases. In some temperature ranges, e.g., $T=278.15$K--$T=298.15$K,
the maximum reduction volume fraction actually increases as pressure
increases, and the excess
enthalpy becomes more negative as the
pressure increases \cite{TomaszkiewiczTA1986II,TomaszkiewiczTA1986III}.

In the dimer model, when $a_{\rm M}>1.0a$, all the cases can reproduce
qualitatively the enthalpy of mixing rather well. As $a_{\rm M}$ increases,
the excess enthalpy becomes more negative and the maximum reduction
volume fraction becomes smaller, following the same trend as in experiments.  The magnitude of the
excess enthalpy for $a_{\rm M}=1.7a$ and $a_{\rm M}=1.8a$
 can be made close to the experimental results, but for the
best model developed in Section III.A, the agreement is only qualitative.

\begin{figure}[H]
\includegraphics[width=0.5\textwidth,angle=0]{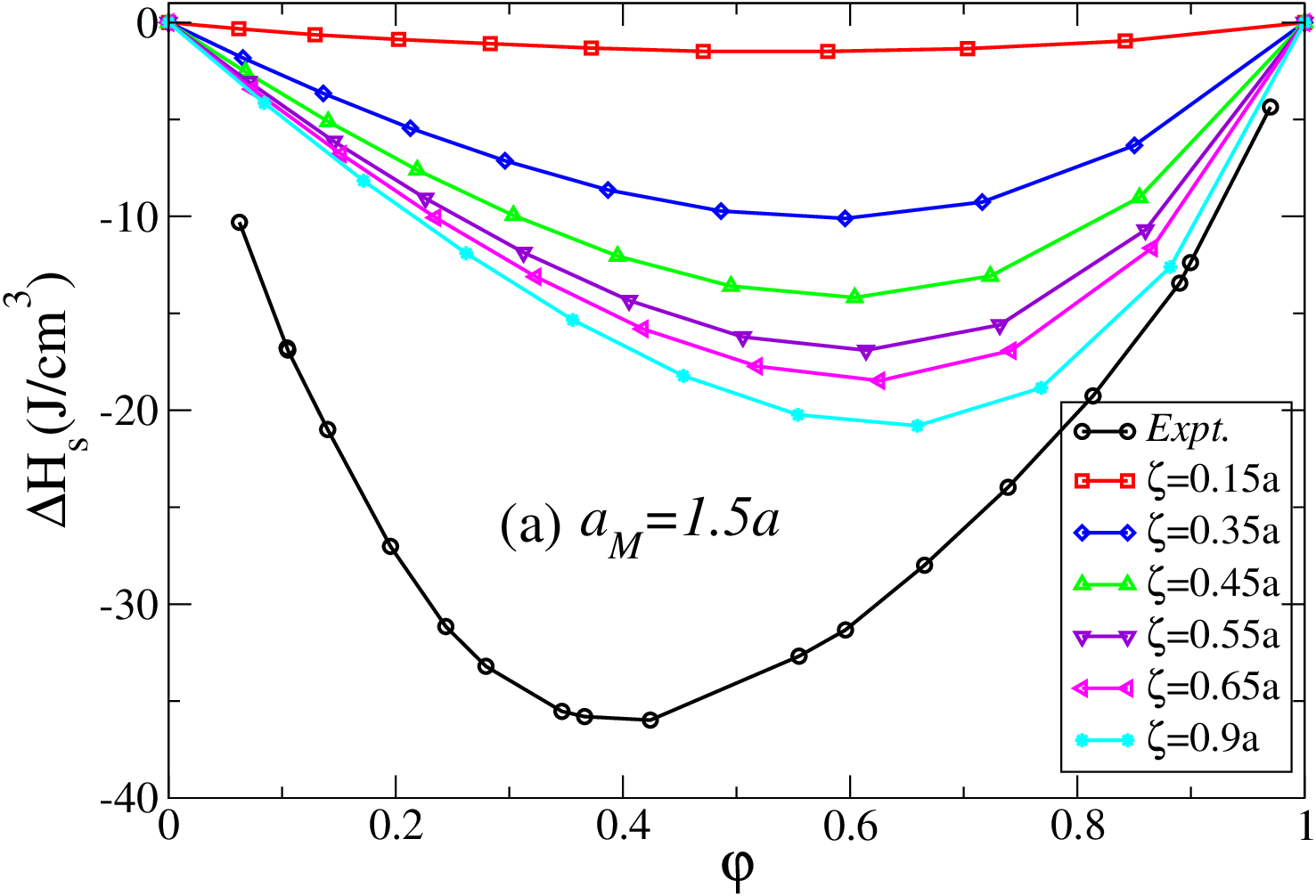}\\
\includegraphics[width=0.5\textwidth,angle=0]{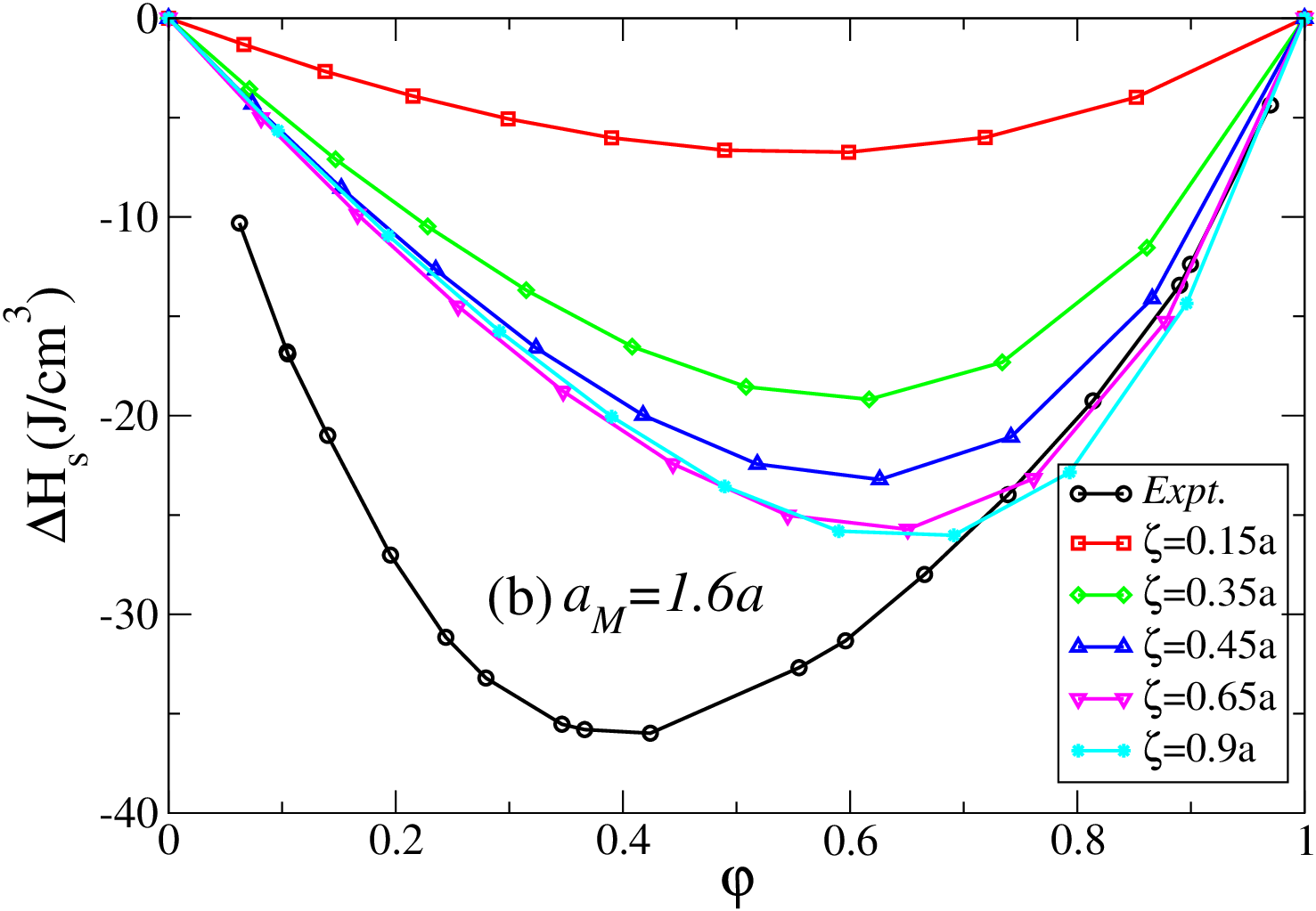} \\
\includegraphics[width=0.5\textwidth,angle=0]{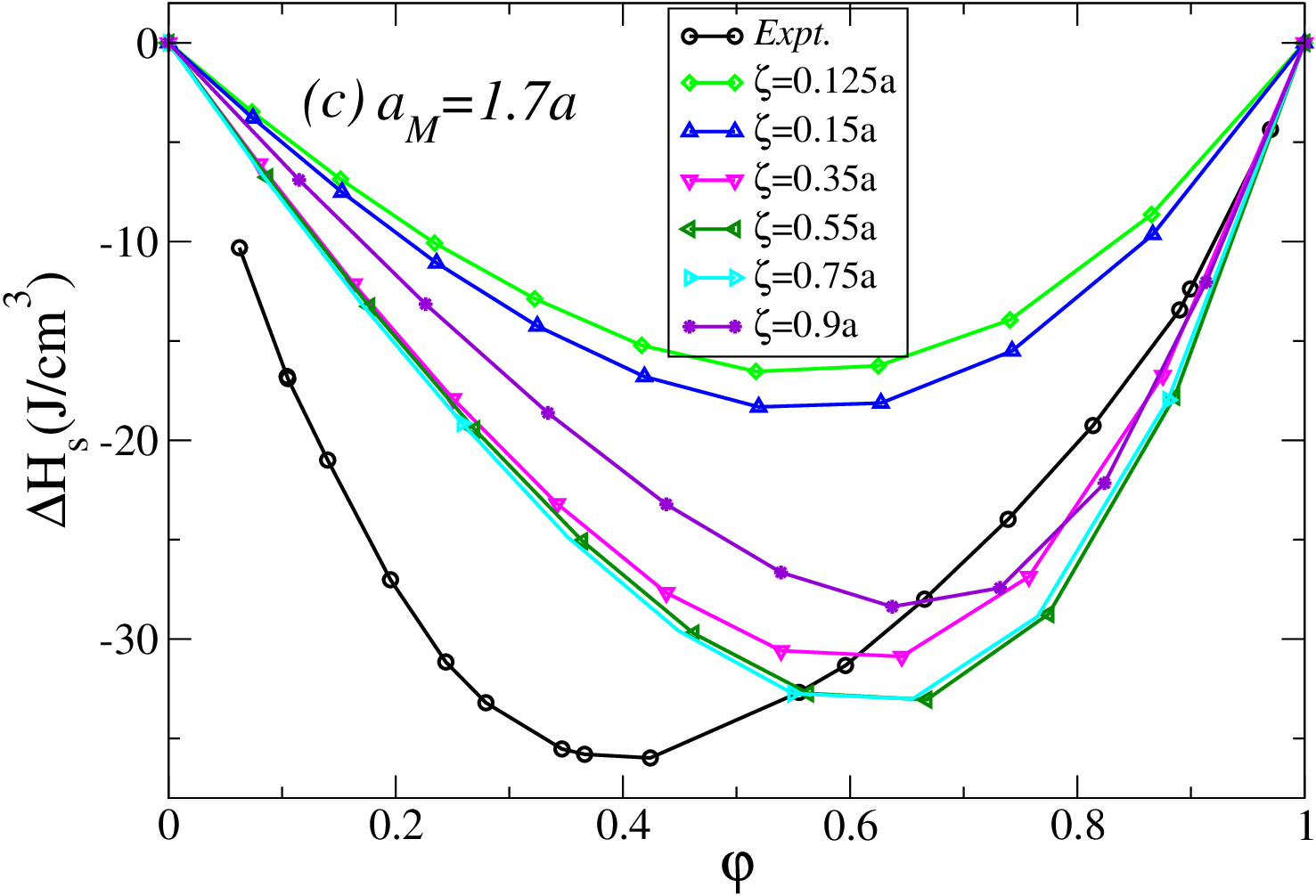}
\caption{(color online) Dependence of the excess enthalpy on solute volume fraction for various values of the bond length, at
  three different hard core diameters $a_{\rm M}$, in the dimer
  model. The simulations are done at $P=0.02$ and $T=0.05$. (a) $a_{\rm
    M}=1.5a$, (b) $a_{\rm M}=1.6a$, (c) $a_{\rm M}=1.7a$. For (a) and
  (b), as the bond length increases, the excess enthalpy becomes more
  negative. For (c), the excess enthalpy exhibits a non-monotonic dependence on bond length. The maximum
  reduction volume fraction increases as $\zeta$ increases for all three
  cases.}
\label{enthalpybonds}
\end{figure}

We report the bond length dependence of the excess enthalpy at $a_{\rm
  HS}=1.5a$ [Fig.~\ref{enthalpybonds}(a)], $a_{\rm M}=1.6a$
[Fig.~\ref{enthalpybonds}(b)], and $a_{\rm M}=1.7a$
[Fig.~\ref{enthalpybonds}(c)].  For $a_{\rm M}=1.5a$ and $a_{\rm
  HS}=1.6a$, as the bond length increases, the excess enthalpy becomes
more negative. For $a_{\rm M}=1.7a$, the excess enthalpy first becomes
more negative and then less negative as the bond length increases.  For
all three values of $a_{\rm M}$, the maximum reduction volume faction
becomes larger as the bond length increases and the maximum enthalpy
reduction is smaller than the experimental value.  For the monomer
model, when $a_{\rm M} > 1.6a$, the excess enthalpy can also be
qualitatively reproduced, and as $a_{\rm M}$ increases it
becomes more negative and the maximum reduction volume fraction
increases.  The value of the excess enthalpy from simulations is smaller
than the experimental values, but the magnitude is comparable.

\begin{figure}[H]
\includegraphics[width=0.5\textwidth,angle=0]{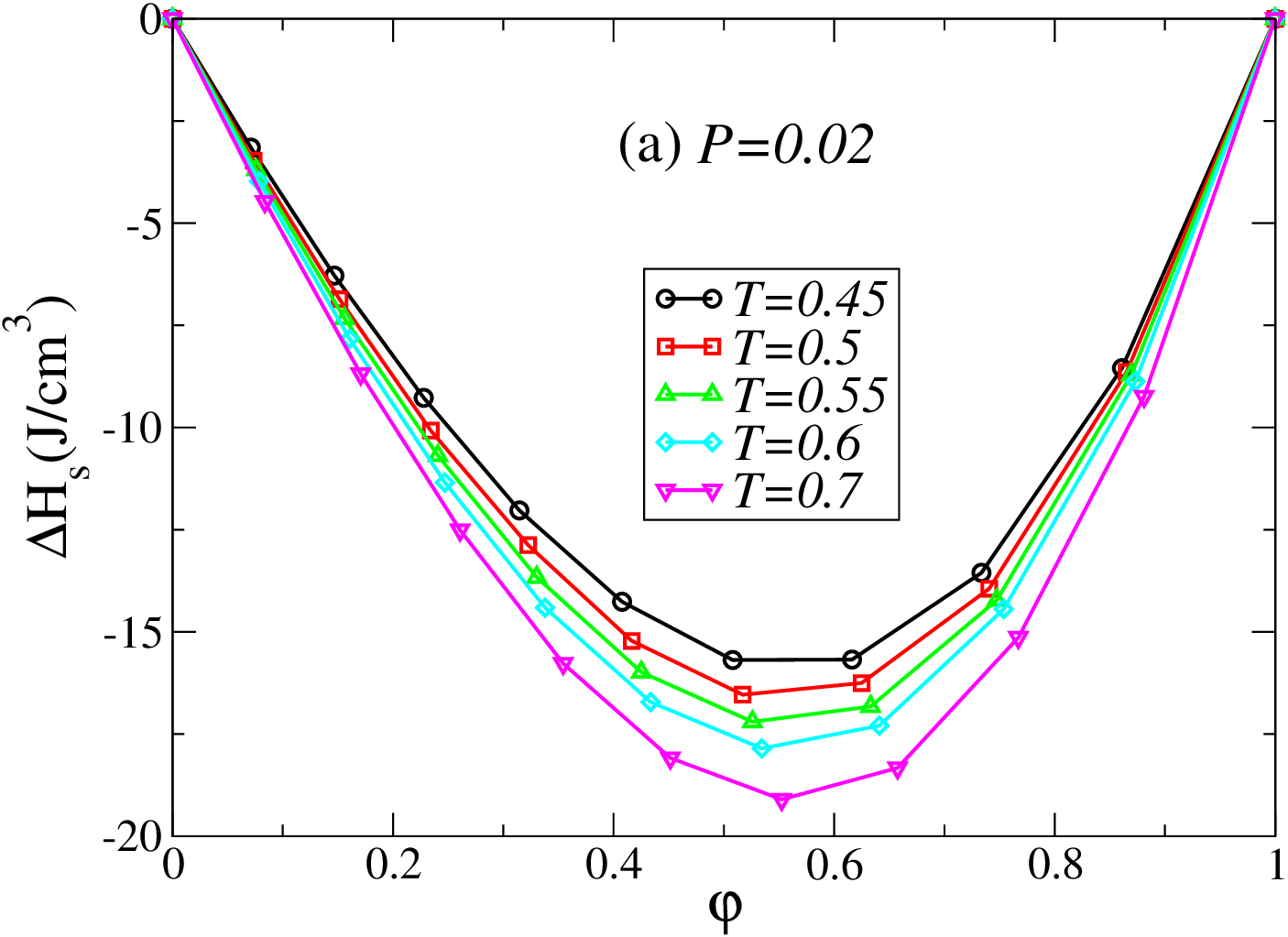} \\
\includegraphics[width=0.5\textwidth,angle=0]{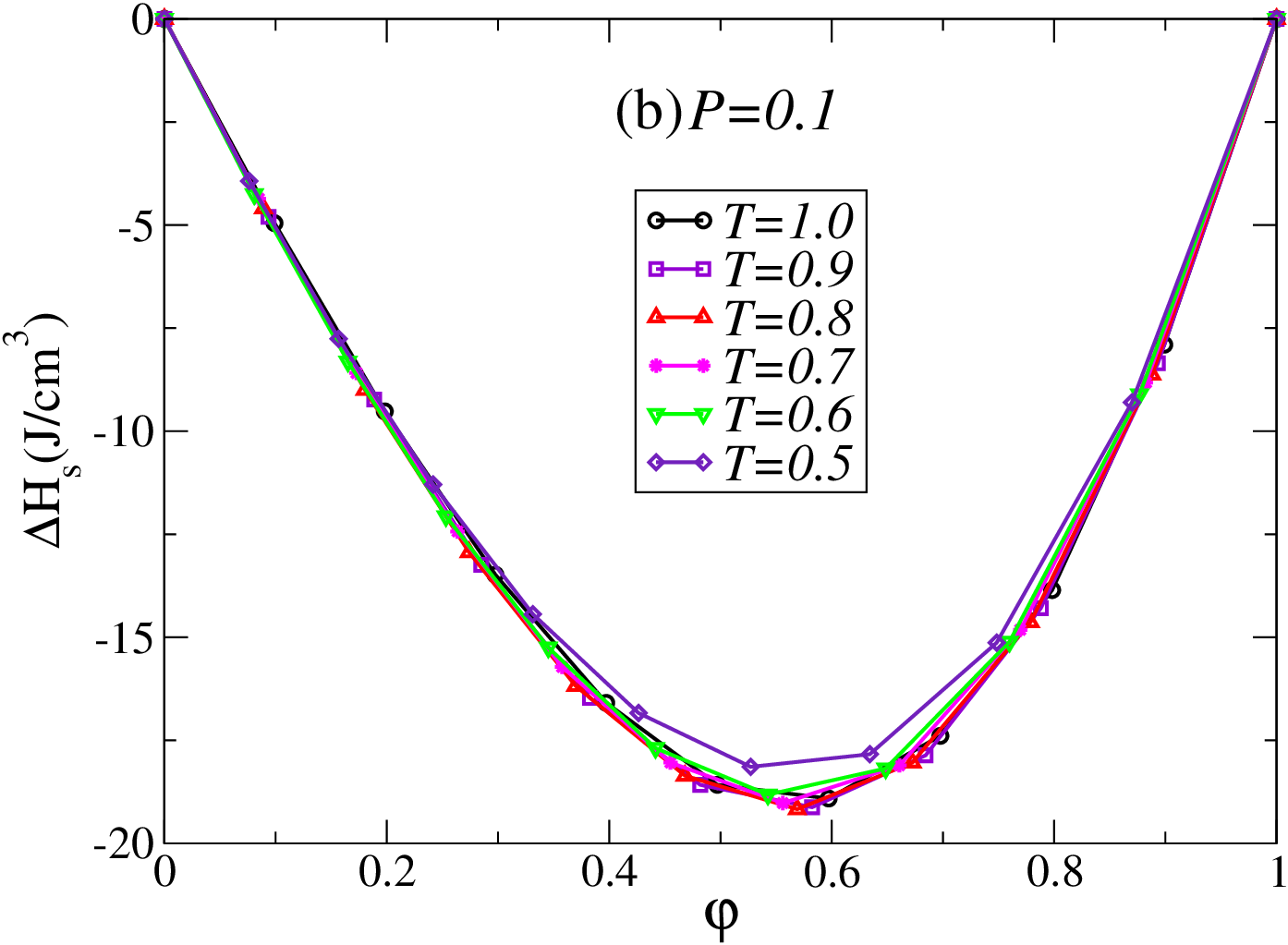}\\
\includegraphics[width=0.5\textwidth,angle=0]{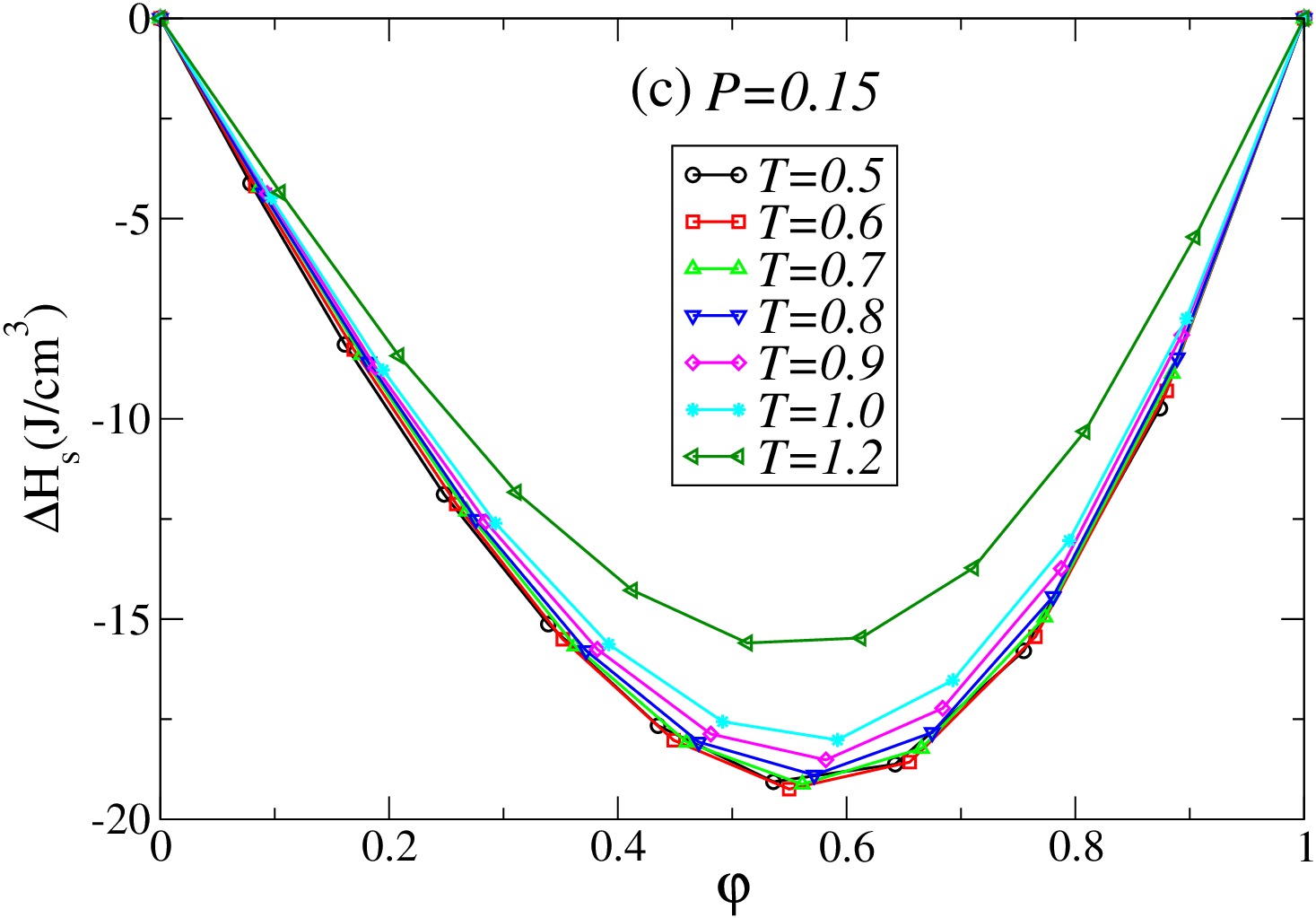}
\caption{(color online) Dependence of the excess enthalpy on solute volume fraction for various temperatures, 
  for the dimer model with $a_{\rm M}=1.7a$, $\zeta=0.125a$.  (a)
  $P=0.02$. As the temperature increases, the excess enthalpy becomes
  more negative, in contrast to experiment, and the maximum reduction
  volume fraction increases \cite{TomaszkiewiczTA1986II}.  (b) $P=0.1$. The excess enthalpy is
  nearly independent of temperature.  (c) $P=0.15$. 
  When $T \geq 0.6$, the excess
  enthalpy decreases as the temperature increases and the volume fraction 
  corresponding to maximum exothermicity shifts towards slightly higher values.}
\label{enthalpytemperature}
\end{figure}

We return to the dimer with $a_{\rm M}=1.7a$ and $\zeta=0.125a$ and
examine the temperature dependence. At $P=0.02$
[Fig.~\ref{enthalpytemperature}(a)], our simulation results contradict
the experimental data: the excess enthalpy becomes increasingly negative
as the temperature increases. However, the increasing maximum reduction
volume fraction with increasing temperature does agree with the experimental trend.  At
$P=0.1$ [Fig.~\ref{enthalpytemperature}(b)], the excess enthalpy changes
only very slightly with temperature, although at the highest $T$ the
magnitude of the excess enthalpy decreases slightly with $T$. If we
increase the pressure to $P=0.15$ [Fig.~\ref{enthalpytemperature}(c)],
we can see that as the temperature increases, the magnitude of excess
enthalpy decreases with $T$ and the maximum reduction volume fraction
increases, in agreement with experimental trends.

\begin{figure}[H]
\centerline{\includegraphics[width=0.8\textwidth,angle=0]{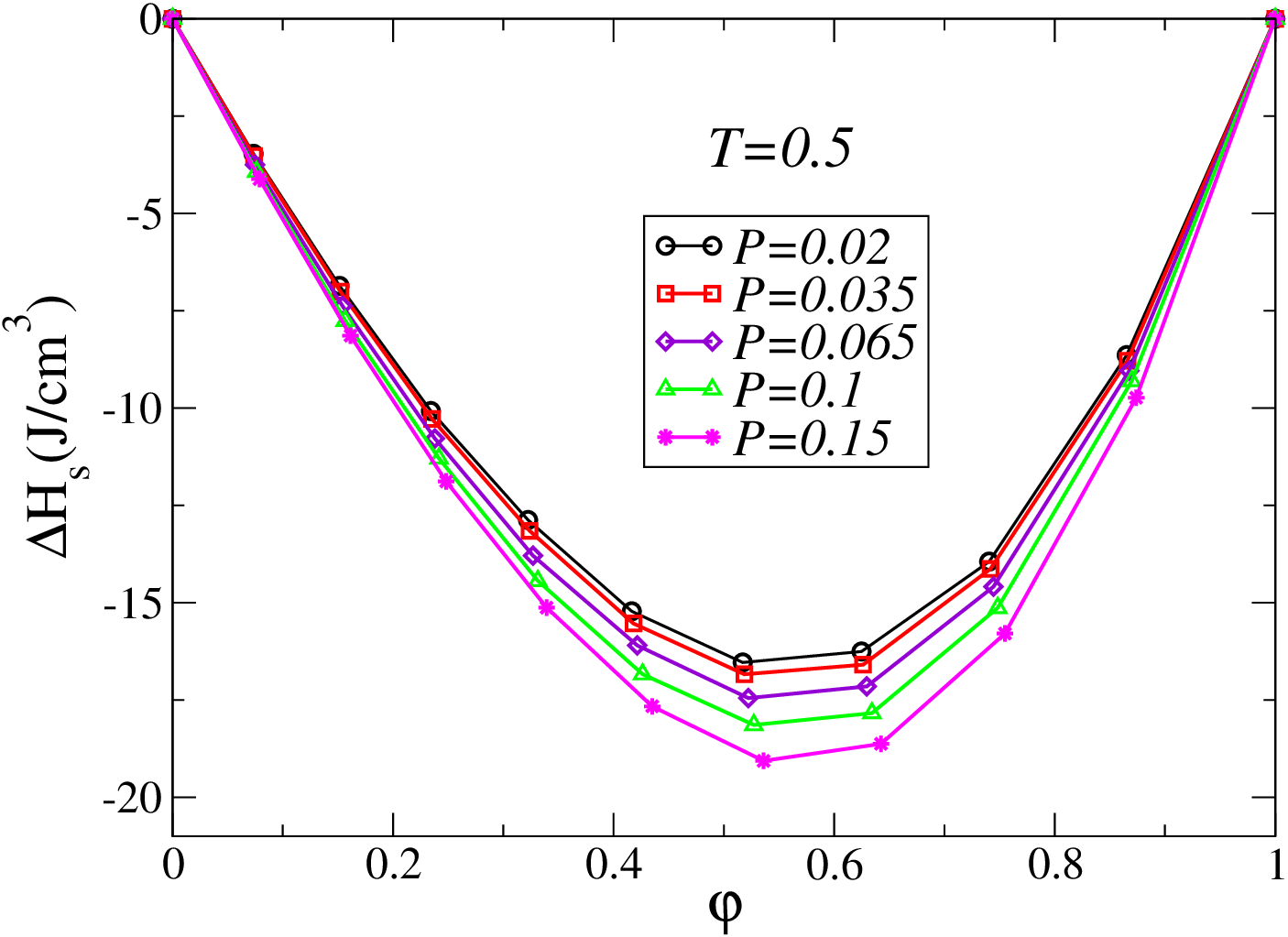}}
\caption{(color online) Dependence of the excess enthalpy on solute volume fraction 
  for various values of the pressure, at $T=0.5$, for the dimer model with $a_{\rm M}=1.7a$,
  $\zeta=0.125a$. As the pressure increases, the excess enthalpy
  increases, and the volume fraction corresponding to maximum exothermicity increases.}
\label{enthalpypressure}
\end{figure}

In our simulation of the pressure dependence of the excess enthalpy at
$T=0.5$ (Fig.~\ref{enthalpypressure}), as the pressure increases, the
excess enthalpy becomes more negative and the maximum reduction volume
fraction becomes larger, which agrees well with experimental
observations of methanol-water solutions.

\subsection{Effect on the temperature of maximum density}
At sufficiently low temperatures and pressures, the density of liquid water
exhibits a maximum with respect to temperature at fixed pressure.
For example, liquid water has a maximum density at
$T=277$K and $P=0.1$MPa. If we add solutes to water, the temperature of
the maximum density changes. According to Ref.~\cite{WadaBCSJ1962}, the
$T_{\rm MD}$ of a methanol solutions reaches its maximum at 
around $x = 0.6\%$ mole fraction and then decreases slightly as the
mole fraction increases, reaching 269K at $x = 5\%$. 
This non-monotonic behavior has been explained by Chatterjee
et al. using a statistical mechanical model of water
\cite{DebenedettiJCP2005}. 

\begin{figure}[H]
\includegraphics[width=0.7\textwidth,angle=0]{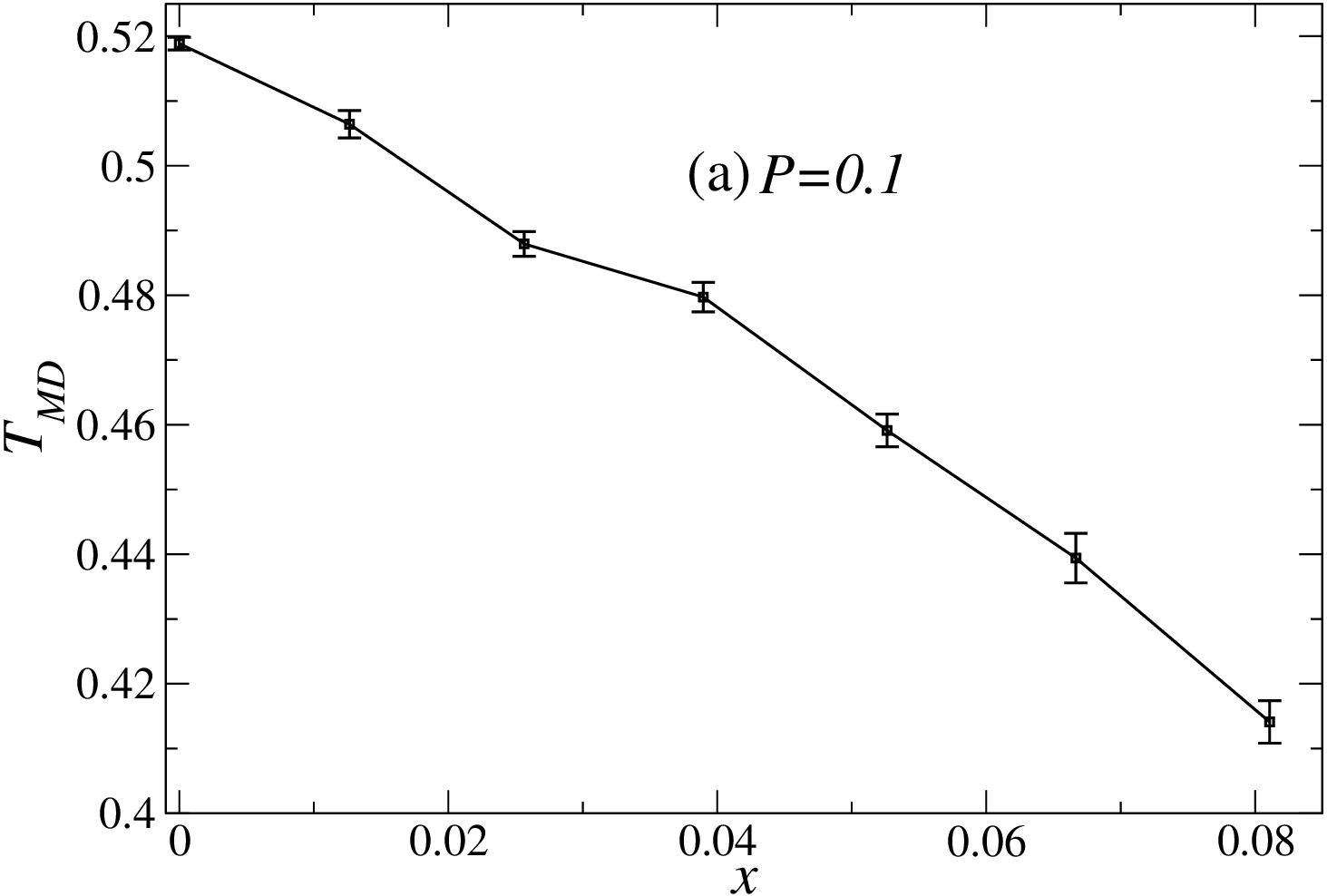}\\
\includegraphics[width=0.7\textwidth,angle=0]{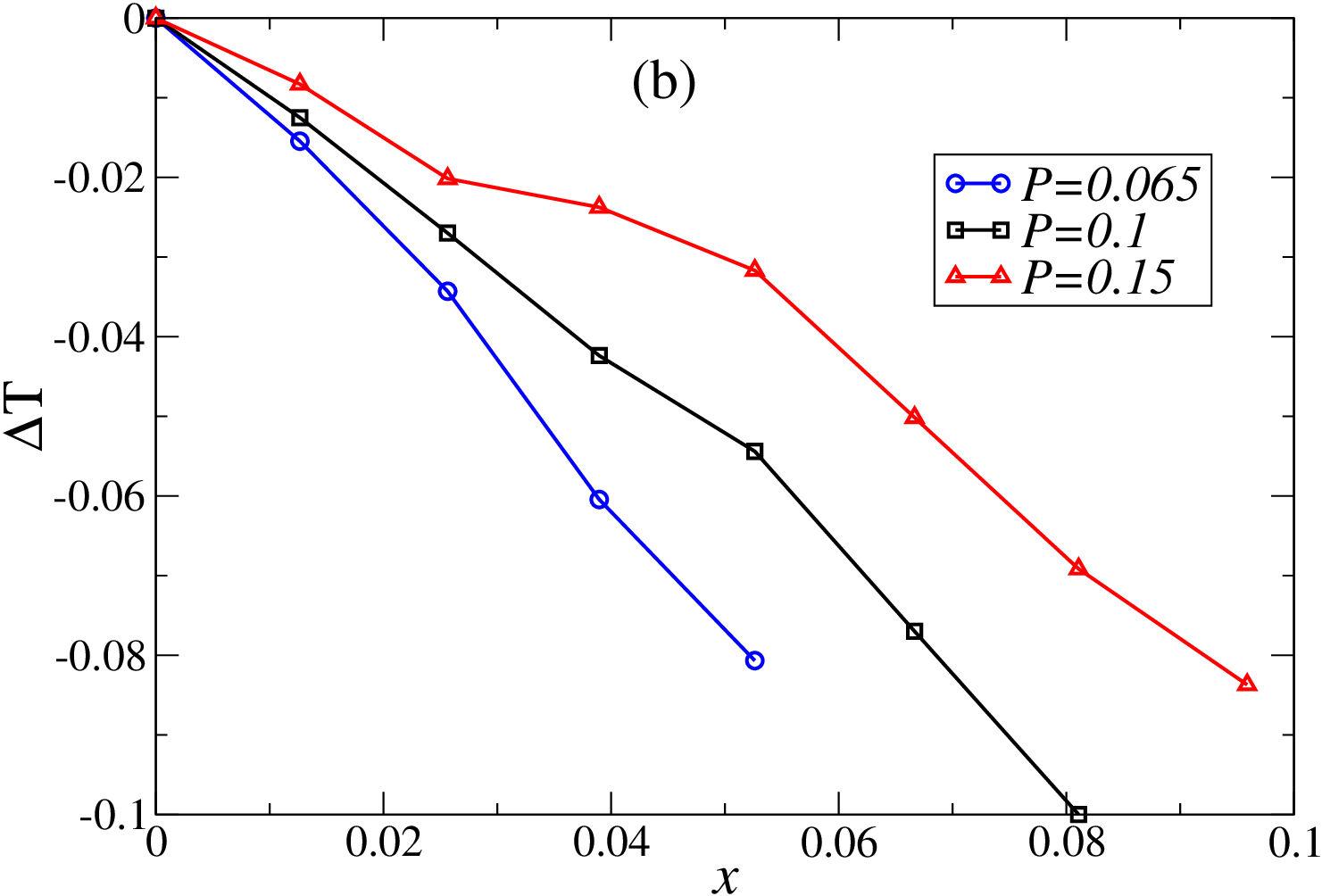}
\caption{(color online) Effect of amphiphilic solutes on $T_{\rm
    MD}$ of solutions.  We use the dimer with parameters of $a_{\rm
    M}=1.7a$, $\zeta=0.125a$. (a) The $T_{\rm MD}$ of the solutions
  at different solute
  concentrations, for $P=0.1$. (b) The shift of $T_{\rm MD}$, 
  with respect to that of the pure solvent,
  $\bigtriangleup T$, as a function of solute mole fraction, $x$, 
  at three different pressures.  Note that $\bigtriangleup T$ is always negative, 
  indicating that adding solute lowers the solution's $T_{\rm MD}$ 
  (lower temperatures are needed to observe negative thermal expansion). For a given mixture, the magnitude
  of $\bigtriangleup T$ decreases as the pressure increases.}
\label{TMD}
\end{figure}

We explore the change of the $T_{\rm MD}$ with concentration for our model. We define the change of the
temperature of maximum density as $\bigtriangleup T = T_{\rm MDs} -
T_{\rm MDJ}$ where $T_{\rm MDs}$ is $T_{\rm MD}$ of the solution and
$T_{\rm MDJ}$ is the $T_{\rm MD}$ of the pure Jagla liquid, at the given pressure.
We have carried out simulations at three different pressures 
for the case of dimer with $a_{\rm M}=1.7a$ and $\zeta=0.125a$.
Our results are shown in Table I and Fig.~\ref{TMD}. 
We find that $\bigtriangleup T$ is always
negative and its absolute value increases monotonically with solute mole fraction.  
If we increase pressure for the same mole fraction, the
change of $T_{\rm MD}$ decreases in magnitude, but it never becomes
positive. Moreover, we find that the decrease of $T_{\rm MD}$ in our
model is orders of magnitude stronger in methanol-water mixtures.

\section{CONCLUSION}

Inspired by the distinctive properties of methanol-water solutions, we
construct a dimer with a hard sphere and a Jagla particle to model
amphiphilic solutes. We vary the hard core diameter and the bond length
to achieve the best agreement between
simulations and experiment in the excess volume. We find that the best agreement
occurs for the dimer with $a_{\rm M}=1.7a$ and $\zeta=0.125a$, which
suggests that the dimer model can be reduced to a monomer with a large
hard core and an attractive potential that coincides with the attractive
part of the Jagla potential. Regarding the temperature and
pressure dependence of the excess volume, our results agree qualitatively
with experimental data. Our model reproduces the excess enthalpy
of the methanol solutions less accurately than the excess volume.  This
is related to the fact that, in our simple model of amphiphilic solutes,
we use the unchanged Jagla potential for the amphiphilic group. We
speculate that a better agreement could be achieved if we varied the
potential of the amphiphilic group.  When we investigate the effect of
the amphiphilic solute on the temperature of maximum density of a
solution, we find that unlike in water-methanol solution, the $T_{\rm MD}$ monotonically
decreases with solute concentrations. Moreover, the effect of concentration in the model
is orders of magnitude stronger than in experiments.

\subsection*{Acknowledgments}

ZS and HES thank the NSF Chemistry Division (grant CHE 0908218) for support.
SVB acknowledges the partial support of this research by the Dr. Bernard W. Gamson
Computational Science Center at Yeshiva College. 
PGD and PJR gratefully acknowledge the support of the National Science
Foundation (Collaborative Research Grants CHE-0908265 and
CHE-0910615). PJR also gratefully acknowledges additional support from
the R. A. Welch Foundation (F-0019).

\newpage

\begin{table}
\caption{The temperature of maximum density, $T_{\rm MD}$, and the difference between the $T_{\rm MD}$ 
  of the solution and that of  the pure solvent at the given pressure, $\bigtriangleup T$, for solutions with
  different mole fractions of amphiphilic solutes in Jagla solvents, at
  different pressures. Results are for the dimer solute model with $a_{\rm M}=1.7a$,
  $\zeta=0.125a$.}
\bigskip
\begin{tabular}{|r|rr|rr|rr|}\hline
 & P=0.065&& P=0.1&&P=0.15& \\
x$\times$100 & $T_{\rm MD}$ & $\bigtriangleup T$ & $T_{\rm MD}$ &
$\bigtriangleup T$& $T_{\rm MD}$ & $\bigtriangleup T$ \\ 
\hline
0  &0.507&0&  0.516&0&0.510&0 \\
1.27&0.491&-0.016& 0.504&-0.012&0.502&-0.008 \\
2.56&0.472&-0.035& 0.489 &-0.027&0.490&-0.02\\
3.90&0.446&-0.062&0.474 &-0.04&0.486&-0.024\\
5.26&0.426&-0.81&0.462&-0.51&0.478&-0.032\\
6.67 &--~~~~&--~~~~&0.439&-0.77& 0.46&-0.05\\
8.11&--~~~~&--~~~~&0.416 &-0.1&0.441&-0.069\\
9.59&--~~~~&--~~~~&--~~~~&--~~~~&0.426&-0.0837\\
\hline
\end{tabular}
\end{table}

\newpage

\newpage

\newpage

\newpage

\newpage

%

\newpage

%

\newpage

%


\newpage

%

\newpage

\newpage

\newpage

\end{document}